\documentclass[prd,aps,twocolumn,a4paper,floatfix,showpacs,nofootinbib,superscriptaddress]{revtex4-1}

\usepackage{times}
\usepackage[usenames]{color}
\usepackage{amsfonts}
\usepackage{amsmath}
\usepackage{amssymb}
\usepackage{bm}
\usepackage{dcolumn}
\usepackage{enumerate}
\usepackage{epsfig}
\usepackage{graphicx}
\usepackage{graphics}
\usepackage{gensymb}
\usepackage{multirow}
\usepackage[latin1]{inputenc}
\usepackage{latexsym}
\usepackage{rotating}
\usepackage{hyperref}
\usepackage{nicefrac}
\usepackage[normalem]{ulem}
\usepackage[caption=false]{subfig}
\usepackage{CJK}
\definecolor{orange}{rgb}{0.9,0.5,0}

\newcommand{\<}{\begin{equation}}
\newcommand{\?}{\end{equation}}

\newcommand{\R}{\mathbb{R}}

\makeatletter
\newcommand*{\llll}{%
  \mathrel{%
    \mathpalette\@charFour<%
  }%
}
\newcommand*{\@charFour}[2]{%
  \sbox0{$\m@th#1#2$}%
  \copy0 %
  \kern-.6\wd0 %
  \copy0 %
  \kern-.6\wd0 %
  \copy0 %
  \kern-.6\wd0 %
  \copy0 %
}
\makeatother

\begin{document}
\begin{CJK*}{GB}{}

\title{Distinguishing high-mass binary neutron stars from binary black holes with second- and third-generation gravitational wave observatories}

\author{An Chen}
\affiliation{Department of Physics, Chinese University of Hong Kong, Sha Tin, Hong Kong}
\affiliation{Department of Physics and Astronomy, University College London, Gower Street, London WC1E 6BT, UK}
\author{Nathan~K.~Johnson-McDaniel}
\affiliation{Department of Applied Mathematics and Theoretical Physics, Centre for Mathematical Sciences, University of Cambridge, Wilberforce Road, Cambridge, CB3 0WA, UK}
\author{Tim Dietrich}
\affiliation{Institute for Physics and Astronomy, University of Potsdam, Karl-Liebknecht-Str.~24/25, 14476, Potsdam, Germany}
\affiliation{Nikhef, Science Park, 1098XG Amsterdam, Netherlands}
\author{Reetika Dudi}
\affiliation{Theoretical Physics Institute, University of Jena, 07743 Jena, Germany}
\affiliation{Max Planck Institute for Gravitational Physics (Albert Einstein Institute), Am M{\"u}hlenberg 1, Potsdam 14476, Germany }

\date{\today}

\begin{abstract}

While the gravitational-wave (GW) signal GW170817 was accompanied by a variety of electromagnetic
(EM) counterparts, sufficiently high-mass binary neutron star (BNS) mergers are expected to be unable
to power bright EM counterparts. The putative high-mass binary BNS merger GW190425, for which no confirmed
EM counterpart has been identified, may be an example of such a system.
Since current and future GW detectors are expected to detect many more BNS mergers, 
it is important to understand how well we will be able to distinguish high-mass BNSs 
and low-mass binary black holes (BBHs) solely from their GW signals. 
To do this, we consider the imprint of the tidal deformability of the neutron stars on 
the GW signal for systems undergoing prompt black hole formation after merger. 
We model the BNS signals using hybrid numerical relativity -- tidal effective-one-body waveforms. 
Specifically, we consider a set of five nonspinning equal-mass BNS signals with total masses of $2.7$, $3.0$, $3.2M_\odot$ 
and with three different equations of state, as well as the analogous BBH signals. 
We perform Bayesian parameter estimation on these signals at luminosity distances 
of $40$ and $98$~Mpc in an Advanced LIGO-Advanced Virgo network 
and an Advanced LIGO-Advanced Virgo-KAGRA network 
with sensitivities similar to the third and fourth observing runs (O3 and O4), 
respectively, and at luminosity distances of $369$ and $835$~Mpc in a network of two 
Cosmic Explorers and one Einstein Telescope, with a Cosmic Explorer sensitivity similar to Stage 2. 
Our analysis suggests that we cannot distinguish the signals from high-mass BNSs and BBHs 
at a $90\%$ credible level with the O3-like network even at $40$~Mpc. 
However, we can distinguish all but the most compact BNSs that we consider in our 
study from BBHs at $40$~Mpc at a $\geq 95\%$ credible level using the O4-like network, 
and can even distinguish them at a $>99.2\%$ ($\geq 97\%$) credible level at $369$ ($835$)~Mpc using the 3G network.
Additionally, we present a simple method to compute the leading effect of the Earth's rotation on the
response of a gravitational wave detector in the frequency domain.
\end{abstract}

\maketitle
\end{CJK*}

%%%%%%%%%%%%%%%%%
\section{Introduction}
\label{sec:intro}
%%%%%%%%%%%%%%%%%

The first binary neutron star (BNS) merger signal detected with gravitational waves (GWs), GW170817~\cite{GW170817}, also generated a panoply of electromagnetic (EM) counterparts~\cite{GW170817_MMA}, notably a gamma-ray burst~\cite{GW170817_GR_GRB,Goldstein:2017mmi} and a kilonova~\cite{Arcavi:2017xiz,Coulter:2017wya,Lipunov:2017dwd,
Soares-Santos:2017lru,Tanvir:2017pws,Valenti:2017ngx}.
In the future, with the growing network of ground-based gravitational wave (GW) detectors, one expects that many more BNS coalescences will be observed~\cite{Aasi:2013wya}, particularly when the third generation of GW detectors comes online~\cite{Mills:2017urp}. 
However, a number of future detections might not have detectable EM counterparts,
either due to the large distance to the source, which makes the detection of an EM signal generally very difficult, 
or due to a different merger scenario compared to GW170817. This was already the case for the more massive and
distant BNS GW190425~\cite{GW190425,S190425z_GCNs}, though its poor sky localization made the identification of any EM counterpart
quite difficult. See Ref.~\cite{Coughlin:2019zqi} for a review about constraints on the ejecta mass 
for all O3a BNS and BHNS candidates.\footnote{There are
also other possible counterparts than the kilonova, short gamma-ray-burst (GRB), or GRB afterglows, which 
might give rise to EM signatures before the moment of merger even for very massive systems, see, e.g., Refs.~\cite{Tsang:2011ad,Palenzuela:2013hu,Paschalidis:2018tsa,Carrasco:2020sxg,Most:2020ami,Nathanail:2020fkp}.
Since the existence of such observables is still under debate, we will not consider these types of scenarios here.}

In general, some BNS systems, as GW190425, will have large enough masses and will be close enough to equal mass so that they will directly collapse to a black hole with negligible matter remaining outside (see, e.g., the discussion in Refs.~\cite{Shibata:2019wef,Coughlin:2018fis,Kiuchi:2019lls}). 
In these cases only a very faint EM counterpart, if any, would be created. 
Of the observed BNS systems in the Milky Way, three have total masses slightly greater than $2.8M_\odot$ (see, e.g., Table~1 in Ref.~\cite{Farrow:2019xnc}), all of which will merge within a Hubble time. For sufficiently soft equations of state (EOSs) of nuclear matter, these could collapse directly to a black hole when they merge; see, e.g., Table~III in Ref.~\cite{Agathos:2019sah}, which finds a threshold mass $\lesssim 2.8M_\odot$ for some EOSs that have maximum masses (of nonrotating stable stars) $\gtrsim 2M_\odot$.\footnote{However, one of these systems, J1913+1102, has a mass ratio $\sim 0.8$, at the approximate threshold for the absence of a significant disk around the final black hole identified in Ref.~\cite{Shibata:2019wef}.}
The EOSs with a low threshold mass are the soft EOSs favored by observations of GW170817's tidal deformability~\cite{GW170817_EOS,GWTC-1,GW170817_model_comp}. The recent population synthesis calculations in Ref.~\cite{Chattopadhyay:2019xye} also predict that more than $30\%$ of merging BNSs observed in GW will have masses $> 3M_\odot$ (weighting by effective volume)---see their Fig.~14.

While one does not expect there to be black holes formed in stellar collapse with masses below the maximum mass of a neutron star, it is possible that primordial black holes could have such masses---see, e.g., Refs.~\cite{Byrnes:2018clq,Carr:2019kxo}---and that they form binaries that merge frequently enough to contribute significantly to the rate of detections made by ground-based detectors---see, e.g., Refs.~\cite{Bird:2016dcv,Clesse:2016vqa,Sasaki:2016jop,Ballesteros:2018swv,Carr:2019kxo,Vaskonen:2019jpv,Young:2019gfc}. Additionally, even without assuming primordial black holes, if the possible mass gap between neutron stars and black holes (see, e.g., Refs.~\cite{Bailyn:1997xt,Ozel:2010su,Farr:2010tu}) does not exist, there could be black holes formed in supernovae with masses just above the maximum mass of a neutron star. For instance, Ref.~\cite{Ertl:2019zks} finds that a small number of such low-mass black holes can be formed via fallback accretion. We thus want to determine how well we can expect to distinguish high-mass BNSs from low-mass binary black holes (BBHs) from their GW signal alone; 
see also Ref.~\cite{Tsokaros:2019lnx} for a recent numerical relativity study on this subject. 
Here we will use the effect of the neutron stars' tidal deformabilities on the signal~\cite{Flanagan:2007ix} to distinguish BNSs from BBHs: Neutron stars have nonzero tidal deformabilities\footnote{In fact, neutron stars have dimensionless tidal deformabilities $> 3$, from causality---see Fig.~3 in~\cite{Zhao:2018nyf}; see also Fig.~1 in~\cite{GW170817_model_comp} for results for a variety of tabulated EOSs.} while black holes have zero tidal deformability (at least perturbatively in the spin of the black hole as well as in the convention used to create the waveform models we use---cf.\ Ref.~\cite{Gralla:2017djj}); see, e.g., Refs.~\cite{Binnington:2009bb,Gurlebeck:2015xpa,Landry:2015zfa,Pani:2015nua}. 
This problem was first studied (without considering high-mass BNSs in particular) in Refs.~\cite{Read:2013zra,Hotokezaka:2016bzh} which applied a simple distinguishability criterion based on the noise-weighted inner product to numerical relativity waveforms. Our study is the first one to investigate this with the Bayesian analysis tools that are applied to real GW data, in addition to focusing on the high-mass BNS case and using hybrid waveforms covering the entire frequency band of the detectors.

We do not consider neutron star-black hole (NSBH) binaries in detail here, since there do not yet exist simulations of NSBH analogues to the BNS systems we consider. However, equal-mass NSBHs with a high-mass neutron star are also expected to have minimal matter remaining outside. For instance, the fit for the final disk mass in Ref.~\cite{Foucart:2018rjc} predicts that an equal-mass NSBH with a nonspinning black hole and a neutron star with compactness greater than $0.195$ (a bit more compact than the second most compact neutron star we consider here) will have no matter outside the final black hole. Moreover, the recent numerical simulations of equal-mass NSBH systems with less compact neutron stars in Ref.~\cite{Foucart:2019bxj} have found that this fit overestimates the amount of matter outside the remnant. We thus give some simple estimates of how well we will be able to distinguish these high-mass BNSs from NSBHs, assuming that we know the EOS with good accuracy, which is likely a good assumption for the 3G observations we are considering, which will take place in $\gtrsim 2044$~\cite{Reitze:2019iox}. We will consider distinguishing high-mass BNSs from NSBHs in detail without such assumptions in future work---this has already been studied from different points of view (without focusing on systems with high-mass neutron stars in particular) in Refs.~\cite{Yang:2017gfb,Hinderer:2018pei,Chen:2019aiw,Coughlin:2019kqf,GW170817_model_comp,Barbieri:2019bdq,Kyutoku:2020xka,Han:2020qmn}.

In this article, we study how well one can estimate the effective tidal deformability of high-mass BNSs and BBH systems with the same masses with observations with three GW detector networks: 
\begin{enumerate}[(i)]
 \item the two Advanced LIGO detectors~\cite{AdvLIGO} and Advanced Virgo~\cite{AdvVirgo}, with sensitivities similar to those during their third observing run (O3);
 \item the two Advanced LIGO detectors, Advanced Virgo, and KAGRA~\cite{KAGRA}, with sensitivities corresponding to optimistic expectations for the fourth observing run (O4);
 \item the proposed third-generation (3G) Einstein Telescope~\cite{Hild:2010id,ET_design} and Cosmic Explorer~\cite{CE,Reitze:2019iox} detectors, with one Einstein Telescope and two Cosmic Explorers (a configuration considered in other studies, e.g., Refs.~\cite{Vitale:2016icu,Hall:2019xmm,Sathyaprakash:2019rom}), and the Cosmic Explorer detectors at a sensitivity similar to Stage~2.
\end{enumerate}
We use waveforms from numerical simulations of high-mass BNS~\cite{Bernuzzi:2014owa,Dietrich:2018phi}, hybridized with the TEOBResumS waveform model~\cite{Nagar:2018zoe} for the lower frequencies not covered by the numerical simulation, and TEOBResumS for the BBH signals. We project these gravitational waveforms onto the three networks and use Bayesian parameter estimation to infer the parameters of the binary that produced the signals,
employing the fast {IMRPhenomPv2\_NRTidal} waveform model~\cite{Dietrich:2018uni}, which is commonly used in the analysis of BNS signals in LIGO and Virgo data (e.g., Refs.~\cite{GW170817_PE,GW170817_EOS,GW170817_testing_GR,GWTC-1,GW170817_model_comp,GW190425}).\footnote{
Since the updated NRTidal approximant {IMRPhenomPv2\_NRTidalv2}~\cite{Dietrich:2019kaq} 
was not available at the start of the project, we restrict our analysis to the original NRTidal model.}
We describe the numerical simulations in Sec.~\ref{sec:nr} and the construction of hybrid waveforms in Sec.~\ref{sec:hybrid}. We then discuss the injections in Sec.~\ref{sec:inj} and the parameter estimation results in Sec.~\ref{sec:results}. Finally, we conclude in Sec.~\ref{sec:conclusions}. We assess the accuracy of the numerical relativity waveforms in Appendix~\ref{app:resolution}, discuss how to include the dominant effect of the Earth's rotation in the frequency-domain detector response in Appendix~\ref{app:time_dep_resp}, and give the prior ranges used in our analyses in Appendix~\ref{app:prior_ranges}.

Unless otherwise stated, we employ geometric
units in which we set $G = c = M_\odot = 1$ throughout the article. 
In some places we give physical units for illustration.
We have used the LALSuite~\cite{LALSuite} and PyCBC~\cite{PyCBC} software packages in this study.

%%%%%%%%%%%%%%%%%
\section{Numerical Relativity Simulations}
\label{sec:nr}
%%%%%%%%%%%%%%%%%

For completeness, we present important details about the numerical relativity 
simulations we employ here; 
see Refs.~\cite{Thierfelder:2011yi,Dietrich:2015iva,Dietrich:2018phi} for more details.

%%%%%%%%%%%%%%%%%
\subsection{Numerical Methods}

The $(3+1)$D numerical relativity simulations presented in this article are 
produced with the BAM code~\cite{Brugmann:2008zz,Thierfelder:2011yi,Dietrich:2015iva} 
and employ the Z4c scheme~\cite{Bernuzzi:2009ex,Hilditch:2012fp} 
for the spacetime evolution and the 1+log and gamma-driver conditions for the 
gauge system~\cite{Bona:1994a,Alcubierre:2002kk,vanMeter:2006vi}.
The equations for general relativistic hydrodynamics are solved in
conservative form with high-resolution-shock-capturing 
technique; see e.g., Ref.~\cite{Thierfelder:2011yi}. 
The local Lax-Friedrichs scheme is used for the numerical
flux computation. The primitive reconstruction is performed with the fifth-order WENOZ
scheme of Refs.~\cite{Borges:2008a,Bernuzzi:2011aq}. 

All configurations have been simulated with three different resolutions 
covering the neutron star diameter with about 64, 96, and 128 points, respectively. 
Most of the simulation data have already been made 
publicly available as a part of the CoRe 
database~\cite{Dietrich:2018phi} (\href{http://www.computational-relativity.org}{\texttt{www.computational-relativity.org}}); 
the additional high resolution data which have been computed for this project
will be made available in the near future. 

Throughout the article, we present results for the highest resolved 
numerical relativity simulation with 128 points. 
In Appendix~\ref{app:resolution},
we present mismatch computation between
target waveforms obtained from different resolutions. 
For additional details about the numerical methods and 
the assessment of resolution uncertainties, 
we refer the reader to Refs.~\cite{Bernuzzi:2011aq,Bernuzzi:2016pie,
Dietrich:2018phi,Dietrich:2018upm}.

%%%%%%%%%%%%%%%%%
\subsection{EOSs employed}

The evolution equations are closed by an EOS connecting the pressure 
$p$ to the specific internal energy $\epsilon$ and rest mass density $\rho$, i.e., 
$p=p(\rho,\epsilon)$.
Here, we use three different zero-temperature nuclear EOSs: ALF2~\cite{Alford:2004pf}, a hybrid EOS with
the variational-method APR EOS for nuclear matter~\cite{Akmal:1998cf} transitioning to color-flavor-locked
quark matter; H4, a relativistic mean-field model with hyperons~\cite{Lackey:2005tk}, based on Ref.~\cite{Glendenning:1984jr};
and 2B, a phenomenological soft EOS from Ref.~\cite{Read:2009yp}.
These EOSs are all modeled by  
piecewise polytropes~\cite{Read:2008iy,Dietrich:2015pxa}. 
Thermal effects are added to the zero-temperature polytropes 
with an additional pressure contribution of the form
$p_{\rm th} = (\Gamma_{\rm th}-1)\rho\epsilon$~\cite{Shibata:2005ss,Bauswein:2010dn}.
Throughout the work, we employ $\Gamma_{\rm th}=1.75$
and present further details about the EOSs in Table~\ref{tab:EOS}.

The 2B EOS, with its maximum mass of $1.78 M_\odot$, is strongly disfavored by observations of $\sim 2M_\odot$ neutron stars~\cite{Antoniadis:2013pzd,Fonseca:2016tux,Arzoumanian:2017puf,Cromartie:2019kug}. We still consider this simulation since it contains the most compact stars for which data have been available for us (compactnesses of $0.205$), and is thus a good proxy for simulations of higher-mass binaries with an EOS that supports a higher mass, which will have high compactnesses. Such simulations are only now possible, due to improvements in initial data construction~\cite{Tichy:2019ouu}, and are still quite preliminary (see also Ref.~\cite{Tsokaros:2019lnx} for simulations of highly compact, very high-mass BNSs with an extreme EOS).

Additionally, the H4 EOS is disfavored by the tidal deformability constraints from GW170817. See Ref.~\cite{GW170817_model_comp} for a direct model selection comparison with other EOSs, and, e.g., Refs.~\cite{Abbott:2018exr,Radice:2018ozg,Coughlin:2018fis,Capano:2019eae} for other studies showing that EOSs like H4 with relatively large radii and tidal deformabilities are disfavored by the GW170817 observation. Nevertheless, because of the limited number of prompt collapse simulations which are available, we still use the H4 EOS simulations to enlarge the BNS parameter space in our study.

\begin{table}[t]
  \centering
  \caption{\label{tab:EOS} Properties of the EOSs used.
    All EOSs use a 4-piece piecewise polytrope, where 
    the outer pieces describes the crust with
    $\kappa_\text{crust}=\kappa_0=8.948185\times10^{-2}$ and
    $\Gamma_\text{crust}=1+1/n_0=1.35692$. The divisions for the
    individual parts are at $\rho_{0,1}={\rho}_\text{crust}$,
    $\rho_{0,2}=8.12123\times 10^{-4}$, and $\rho_{0,3} =
    1.62040 \times 10^{-3}$. The columns 
    refer to: the name of the EOS, the maximum density in the crust,
    the three polytropic exponents $\Gamma_i$
    for the individual pieces, and
    the maximum supported gravitational mass $M_\text{max}$ and maximum
    baryonic mass $M^\text{b}_\text{max}$ of an isolated
    non-rotating star.
    }
  \begin{tabular}{l|cccccc}        
    \hline EOS & ${\rho}_\text{crust}\cdot10^{4}$& $\Gamma_1$ &
      $\Gamma_2$ & $\Gamma_3$ & $M_\text{max}\ [M_\odot]$ & $M^\text{b}_\text{max}\ [M_\odot]$ \\ 
    \hline 
    ALF2 & 3.15606 & 4.070 & 2.411 & 1.890 & 1.99 & 2.32 \\ 
    H4   & 1.43830 & 2.909 & 2.246 & 2.144 & 2.03 & 2.33 \\ 
    2B   & 3.49296 & 2.000 & 3.000 & 3.000 & 1.78 & 2.14 \\
    \hline
  \end{tabular}
\end{table}

%%%%%%%%%%%%%%%%%
\subsection{Configurations}

We consider a total of five nonspinning, equal mass BNS configurations with total masses
($M_\text{tot}=M^A+M^B$) of $2.7$, $3.0$, and $3.2M_\odot$. 
We summarize the binary properties the important binary properties 
in Table~\ref{tab:properties}. We extend the 5 BNS configurations 
with one additional BBH model, for which we do not perform a numerical relativity 
simulation but employ only the effective-one-body approximant which we also employ 
for the construction of the hybrid waveform; see Sec.~\ref{sec:hybrid}.
Since BBH configurations are invariant with respect to their total mass, 
we rescale the equal-mass, nonspinning BBH configuration to match the masses of the BNS 
configurations.

The main information about the internal structure of the stars is 
encoded in their tidal deformabilities $\Lambda^{A,B}$.
The mass-weighted tidal deformability 
introduced by Ref.~\cite{Wade:2014vqa}, based on Ref.~\cite{Flanagan:2007ix},
\begin{equation}
\tilde{\Lambda}=\frac{16}{13} \Lambda^A \left( \frac{M^A}{M_\text{tot}}\right)^4 \left( 12-11\frac{M^A}{M_\text{tot}} \right) + (A \leftrightarrow B) ,
\end{equation}
is a well-measured EOS parameter characterizing 
the leading order tidal contribution to the GW phase. 
Thus, we will focus on $\tilde{\Lambda}$ to determine 
if a BNS and BBH system can be distinguished.

\begin{table*}[t]
  \centering
  \caption{\label{tab:properties} 
  Main properties of the studied configurations. The columns 
  refer to: the simulation name, the simulation's unique identifier in the \texttt{CoRe}
  database, the total mass of the system $M_\text{tot} =M^A+M^B$, the EOS governing the 
  neutron star matter, the mass-weighted tidal deformability $\tilde{\Lambda}$, the compactness
  of one of the stars, $\mathcal{C}^{A,B} = M^{A,B}/R^{A,B}$,
  the prompt collapse mass computed as in Ref.~\cite{Bauswein:2013jpa} $M_{\rm pc}$, the baryonic
  mass remaining outside of the final black hole when the apparent horizon is found $M^\text{b}_\text{rem}$,
  and the dimensionless mass and angular momentum of the final black hole, $M_{\rm BH}/M_\text{tot}$ and $\chi_{\rm BH}$.}
\begin{tabular}{llcccccccc}
     \hline Name & \texttt{CoRe}-ID & $M_\text{tot}$ [$M_\odot$] & EOS & $\tilde{\Lambda}$ & $\mathcal{C}^{A,B}$ & $M_{\rm pc}$ [$M_\odot$] & $M^\text{b}_\text{rem}$ [$M_\odot$] & $M_{\rm BH}/M_\text{tot}$ & $\chi_{\rm BH}$\\
    \hline 
     ${\rm 2B}_{2.7}$   & BAM:0001 & 2.7000 & 2B   & $127$ & $0.205$ & 2.40 & $\lesssim 10^{-4}$ & $0.97$ & $0.78$\\
     ${\rm ALF2}_{3.0}$ & BAM:0011 & 3.0000 & ALF2 & $383$ & $0.178$ & 3.04 & $0.004$ & $0.97$ & $0.82$\\
     ${\rm ALF2}_{3.2}$ & BAM:0016 & 3.2001 & ALF2 & $246$ & $0.191$ & 3.04 & $0.001$ & $0.98$ & $0.81$\\
     ${\rm H4}_{3.0}$   & BAM:0047 & 3.0000 & H4   & $567$ & $0.164$ & 3.20 & $0.07$ & $0.96$ & $0.77$\\
     ${\rm H4}_{3.2}$   & BAM:0052 & 3.2000 & H4   & $359$ & $0.176$ & 3.20 & $0.004$ & $0.98$ & $0.83$\\
    \hline 
    ${\rm BBH}$      & --   & --      & --  & 0 & $0.5$ & 0 & $0$ & $0.95$ & $0.69$\\
    \hline
  \end{tabular}
\end{table*}

%%%%%%%%%%%%%%%%%
\subsection{Dynamical Evolution}

\begin{figure}[tb]
\includegraphics[width=0.53\textwidth]{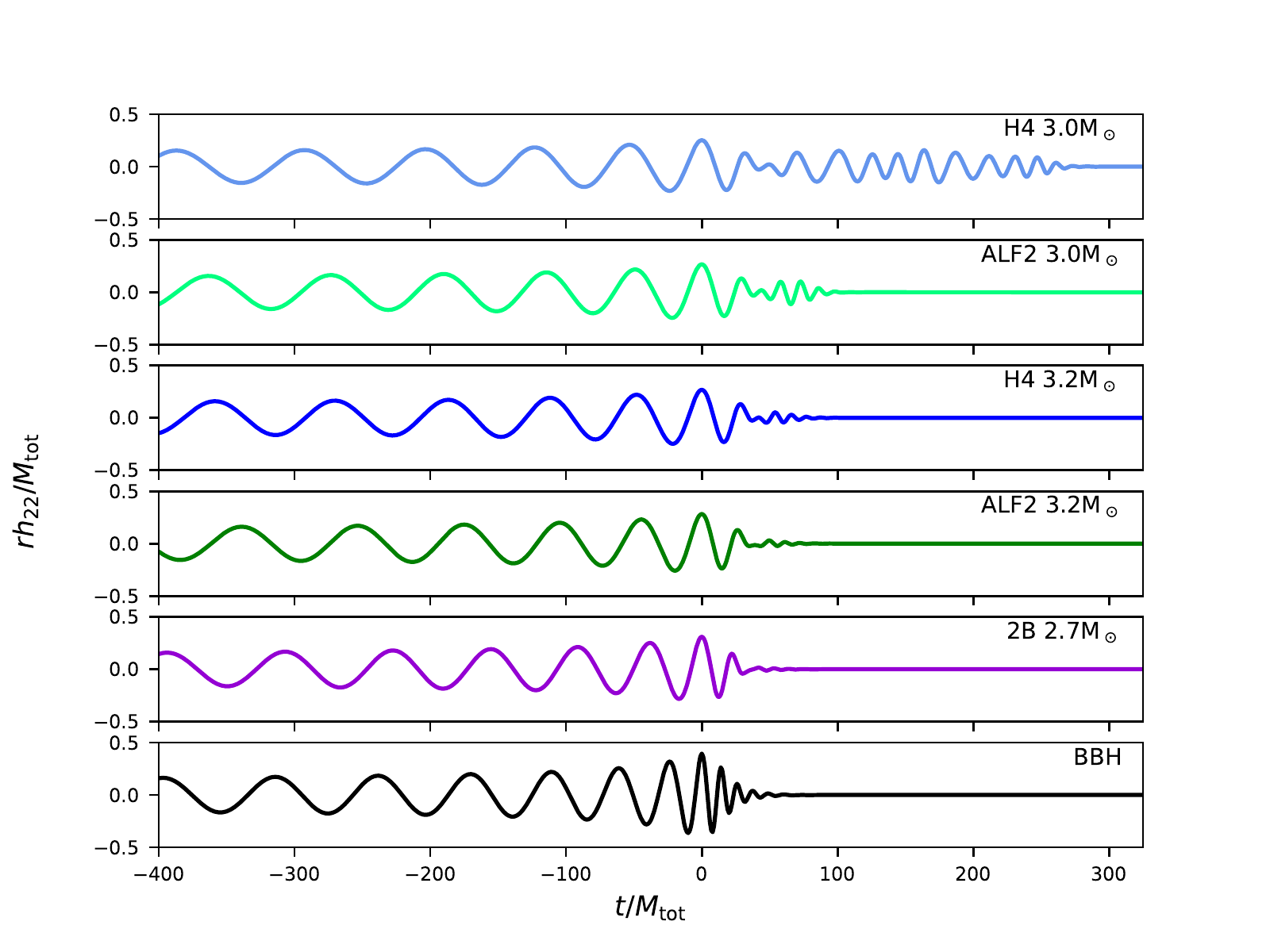}
\caption{The dominant quadrupolar ($\ell = m = 2$) mode of the BNS hybrid waveforms and TEOBResumS BBH waveform around merger versus time $t$, ordered from top to bottom from largest to smallest tidal deformability (cf.\ Table~\ref{tab:properties}). These waveforms are scaled by the distance $r$ and total mass $M_\text{tot}$ and aligned in time and phase, with the peak at $t = 0$, so that they would be identical if they were all BBH waveforms.}
\label{fig:NR_GWs}
\end{figure}

In the following, we give a qualitative overview of the simulation dynamics
and some important simulation results. 
For this, we show for all employed cases the GW signal in Fig.~\ref{fig:NR_GWs}. 
One finds that for a decreasing value of $M_\text{tot}/M_{\rm pc}$, the system systematically
differs from the BBH case, and this difference is imprinted in the waveform. Here $M_{\rm pc}$ is the prompt collapse
mass computed using the expressions from Ref.~\cite{Bauswein:2013jpa}, given in Table~\ref{tab:properties}.
In particular for ${\rm H4}_{3.0}$, we find that after the merger,
the GW signals contains a short postmerger evolution for about $200M_\text{tot}$, which corresponds 
to about $3$~ms. For all other setups, the postmerger GW signal is significantly shorter
or is even missing completely, e.g., ${\rm 2B}_{2.7}$.

As discussed in the introduction, an important reason for our study is the likely scenario in which a
prompt collapse configuration ($M_\text{tot}>M_{\rm pc}$) generates EM signatures which are too faint to
be detected. To a good approximation, the luminosity of the EM counterpart is connected to the 
amount of ejected material which remains outside the black hole---see, e.g., Ref.~\cite{Grossman:2013lqa}.
Thus, we compare the amount of baryonic mass in our 
employed dataset and report in Table~\ref{tab:properties} 
the remnant baryonic mass outside of black hole.
This mass estimate can be seen as an upper bound on the possible ejecta mass, 
since it refers to the total baryonic mass outside the horizon (including bound and unbound material). 
Within our set of simulations, ${\rm H4}_{3.0}$ has more than an order of magnitude more remnant baryonic mass than all other configurations.
As expected, this is also the simulation with the longest postmerger signature.
For comparison, according to~\cite{Coughlin:2018fis}, the total ejecta mass for GW170817 is about $0.05 M_\odot$
and therefore much larger than the total remnant mass of any of our configurations, except ${\rm H4}_{3.0}$.

We also report the dimensionless mass and angular momentum of the final BH formed in the coalescence in Table~\ref{tab:properties}, for comparison with the BBH values. 
We point out that, interestingly, for the BBH setup the final BH mass $M_{\rm BH}/M_\text{tot}$ is a smaller fraction of the initial mass of the system than for the BNS configurations, and the final spin is also smaller in the BBH case. 
This might be caused by the fact that during the BNS coalescence, 
the neutron stars with their larger radii come into contact earlier than for the corresponding
BBH configuration, thus, the emission of GWs is suppressed and less energy and angular momentum is radiated. 

%%%%%%%%%%%%%%%%%
\section{Construction of hybrid waveforms}
\label{sec:hybrid}
%%%%%%%%%%%%%%%%%

While the numerical relativity simulations discussed in the previous
section allow us to model the last stages of the binary
neutron star coalescence, they are orders of magnitude too short ($\sim 0.1$~s) to provide the entire signal in the band
of an interferometeric gravitational wave detector (which can last for minutes or longer).
Thus, in order to model the expected detector response, we have to combine the numerical relativity waveforms with the
predictions of a waveform model at lower frequencies.

For this purpose, we hybridize the NR waveforms with waveforms generated using the TEOBResumS~\cite{Nagar:2018zoe} model. TEOBResumS is an effective-one-body (EOB) model, 
which is based upon a resummation of the post-Newtonian description of the two body problem, 
with further input from NR data, to provide a reliable description in the strong gravity regime, i.e., close to merger. 
For BNSs, tidal effects are incorporated by computing a resummed attractive potential~\cite{Bernuzzi:2014owa}. 
For BBHs, a phenomenological merger-ringdown model tuned to NR simulations is attached to the end of the EOB evolution.

Each waveform is determined by the binary mass-ratio $q$, the total mass $M_\text{tot}$, the multipolar tidal deformability parameters for the two stars $\Lambda_\ell^{A, B}$ ($\ell \in\{2,3,4\}$), and the dimensionless spins $\chi^{A,B}$. 
TEOBResumS incorporates spin-orbit coupling up to next-to-next-to-leading order and the EOS-dependent self-spin effects (or quadrupole-monopole term) up to next-to-leading order. 
In this work, we focus on the $\ell = |m| = 2$ modes of the GW signal and neglect the influence 
of higher modes, which however contribute only about $\lesssim 1\%$ to the total energy budget for equal mass mergers; cf.\ Fig.\ 16 of Ref.~\cite{Dietrich:2016hky}. Additionally, we calculated the SNR of the $(\ell, |m|) = (3, 2), (4, 4)$ modes individually (the dominant higher modes for an equal mass, nonspinning system) using the IMRPhenomHM BBH waveform model~\cite{London:2017bcn} and found that they only have SNRs of $\sim 0.3$ and $\sim 0.6$, respectively for the most optimistic case (the 3G network, with a distance of $369$~Mpc). Thus, the lack of higher modes in the injections is not expected to significantly affect parameter estimation results in these cases.

\begin{table*}
  \centering
  \caption{\label{tab:SNRs} Optimal SNRs of the injections we consider, starting from $31$~Hz (the low-frequency cutoff used in our parameter estimation study), with the SNRs from the detectors' fiducial low-frequency cutoffs given in parentheses, for comparison. We round to the nearest integer and only give the dependence on mass and distance,
  not EOS or BNS v.\ BBH, since those differences are at most $\pm 1$ due to rounding. 
  }
  \begin{tabular}{cccccccccc}        
    \hline 
    \multirow{3}{*}{$M_\text{tot}$ [$M_\odot$]} & \hphantom{X} & \multicolumn{8}{c}{Network \& Distance [Mpc]}\\
    \cline{3-10}
     & & \multicolumn{2}{c}{O3-like} & \hphantom{X}& \multicolumn{2}{c}{O4-like} &\hphantom{X} & \multicolumn{2}{c}{3G}\\
     \cline{3-4}
     \cline{6-7}
     \cline{9-10}
     & & $40$ & $98$ & & $40$ & $98$ & & $369$ & $835$\\
    \hline 
    $2.7$ & & $41$ ($43$) & $17$ ($18$) & & $71$ ($74$) & $29$ ($30$) & & $215$ ($337$) & $103$ ($162$)\\
    $3.0$ & & $45$ ($47$) & $18$ ($19$) & & $77$ ($81$) & $32$ ($33$) & & $234$ ($368$) & $112$ ($176$)\\
    $3.2$ & & $47$ ($49$) & $19$ ($20$) & & $82$ ($85$) & $33$ ($35$) & & $247$ ($388$) & $118$ ($186$)\\
    \hline 
  \end{tabular}
\end{table*}

While GW170817 was detectable from a frequency of $23$~Hz onwards~\cite{GW170817_PE} and one expects to detect BNS signals from even lower frequencies as GW detectors improve, going down to $\sim 5$~Hz for 3G detectors (see, e.g., Ref.~\cite{Meacher:2015rex,Reitze:2019iox}), tidal effects are extremely small in the early inspiral. Therefore, distinguishing BBH and BNS systems at such low frequencies seems impossible. Thus, to reduce the computational expense of this analysis, allowing us to explore more detector networks, BNS configurations, and distances, we restrict our analysis to a frequency interval starting at $31$~Hz (taking a low-frequency cutoff a little above the lowest frequency in the hybrid). We compare the SNRs from $31$~Hz with the SNRs from the fiducial low-frequency cutoffs of the detectors in Table~\ref{tab:SNRs}, as we expect that the information contained in the low-frequency portions of the signal will help us to constrain the binary's masses, spins, and sky position more precisely---see, e.g., Fig.~2 in Ref.~\cite{Harry:2018hke}.

For the construction of the hybrid waveforms, we follow Refs.~\cite{Dudi:2018jzn, Dietrich:2018uni}, which provide further details. We align the NR and TEOBResumS waveforms over a frequency window $[450, 720]$~Hz by minimizing the phase difference. After alignment, we transition between the waveforms by applying a Hann-window function.
For the BBH system, we do not produce a hybrid waveform, and use the BBH-EOB waveform directly, which is a very good representation of full NR results at the given mass ratio of $q=1$. 

%%%%%%%%%%%%%%%%%
\section{Injections and parameter estimation methods}
\label{sec:inj}
%%%%%%%%%%%%%%%%%

As outlined in Sec.~\ref{sec:intro}, we consider 3 detector networks in this study: An O3-like network composed of Advanced LIGO and Advanced Virgo with the ``O3low'' noise curves~\cite{Aasi:2013wya};\footnote{In the first phase of O3, the sky-averaged BNS range of LIGO Livingston was $125$--$140$~Mpc, better than the $120$~Mpc given by the LIGO ``O3low'' noise curve, while for LIGO Hanford was somewhat below this, at $102$--$111$~Mpc. The BNS range for Virgo was $43$--$50$~Mpc, somewhat below the $65$~Mpc given by the Virgo ``O3low'' noise curve. The observed sensitivity numbers come from Ref.~\cite{GW190425}; see also Ref.~\cite{GWO_status} for the detectors' current sensitivity.} an O4-like network composed of Advanced LIGO, Advanced Virgo, and KAGRA, with the design sensitivity noise curves for Advanced LIGO and Advanced Virgo (BNS ranges of $175$ and $120$~Mpc, respectively), plus the $128$~Mpc BNS range noise curve for KAGRA~\cite{Aasi:2013wya};\footnote{The Virgo and KAGRA noise curves are on the optimistic side of predictions for O4.} and a third generation (3G) detector network composed of one Einstein Telescope (ET) detector and two Cosmic Explorer (CE) detectors, using the noise curves from Ref.~\cite{Evans:2016mbw} (the ET-D noise curve and the original standard CE noise curve, which is similar to the sensitivity now predicted for Stage~2).\footnote{The CE Stage~1 and~2 noise curves~\cite{Reitze:2019iox,CE_noise_curves} were not available when we started this study, which is why we use the older standard CE noise curve originally given in Ref.~\cite{CE}. The CE Stage~2 noise curve is similar to the broadband noise curve from Ref.~\cite{CE}, so it is flatter than the older standard curve, but with a smaller maximum sensitivity. Thus, even though the low-frequency cutoff for CE Stage~2 extends down to $5$~Hz, we find that the 3G network SNRs quoted in Table~\ref{tab:SNRs} decrease by $\sim 20\%$ ($\sim 10\%$) with a low-frequency cutoff of $31$~Hz ($5$~Hz; comparing with the $10$~Hz CE cutoff results with the other noise curve). However, the increased high-frequency sensitivity of the CE Stage~2 noise curve may offset the decreased SNR when considering detecting tidal effects. We leave a detailed study of the effects of a different CE noise curve on these results for future work.} The makeup of the 3G network is not yet determined, though 1 ET and 2 CEs is a commonly considered configuration, e.g., Refs.~\cite{Vitale:2016icu,Hall:2019xmm,Sathyaprakash:2019rom}. Since the locations and orientations of 3G detectors are currently unknown, we used the locations and orientations of the two LIGO detectors for the two CE detectors, and the Virgo location for the ET detector, with the orientation of ET given in Ref.~\cite{LALDetectors}. These are not likely to be the actual locations of the 3G detectors---see Ref.~\cite{Hall:2019xmm} for some more likely possibilities---but this choice allows us to keep the same extrinsic parameters as the other injections and still have SNRs close to the average over the extrinsic parameters.

\begin{figure}[tb]
\includegraphics[width=0.53\textwidth]{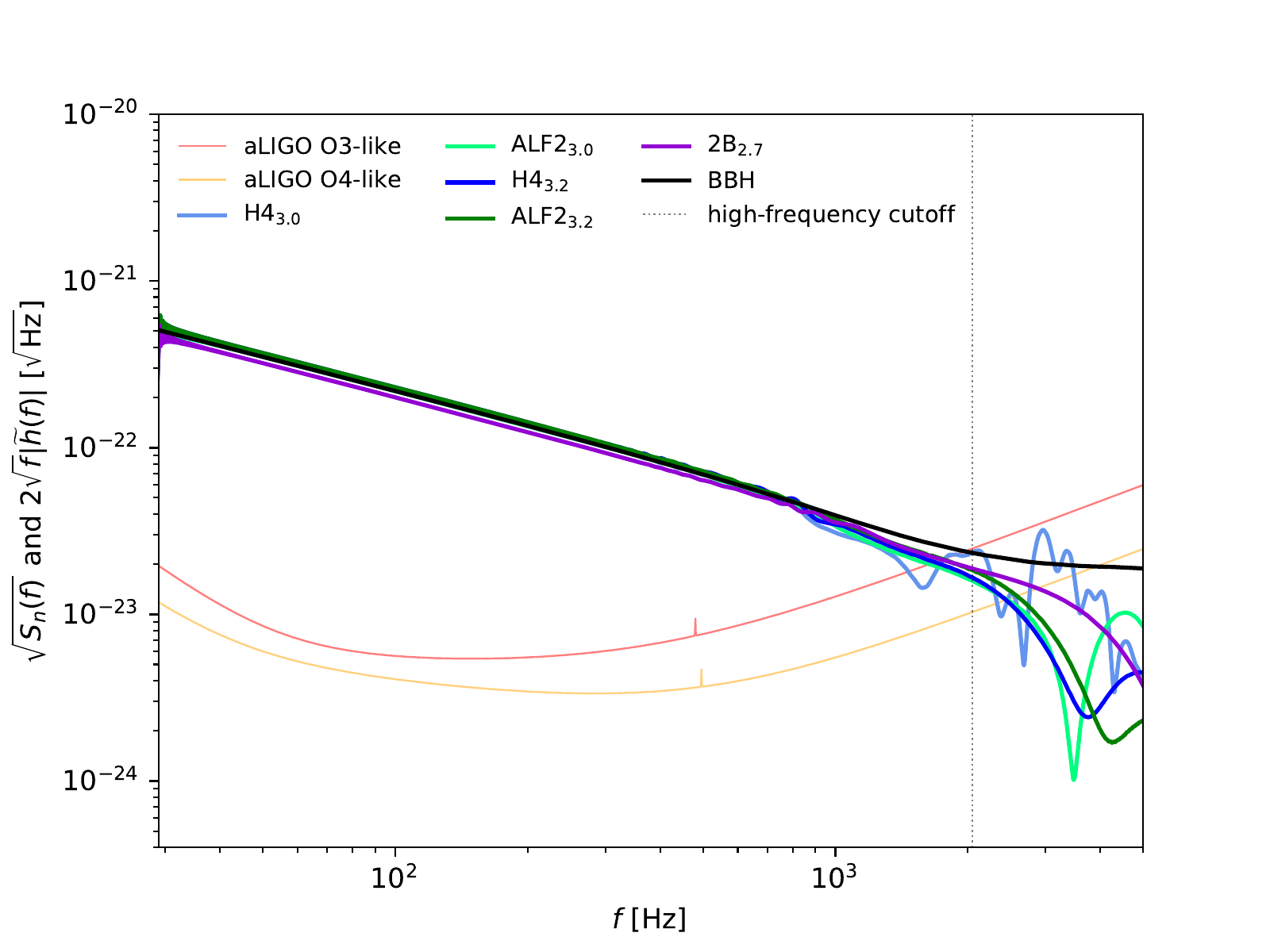}\\
\includegraphics[width=0.53\textwidth]{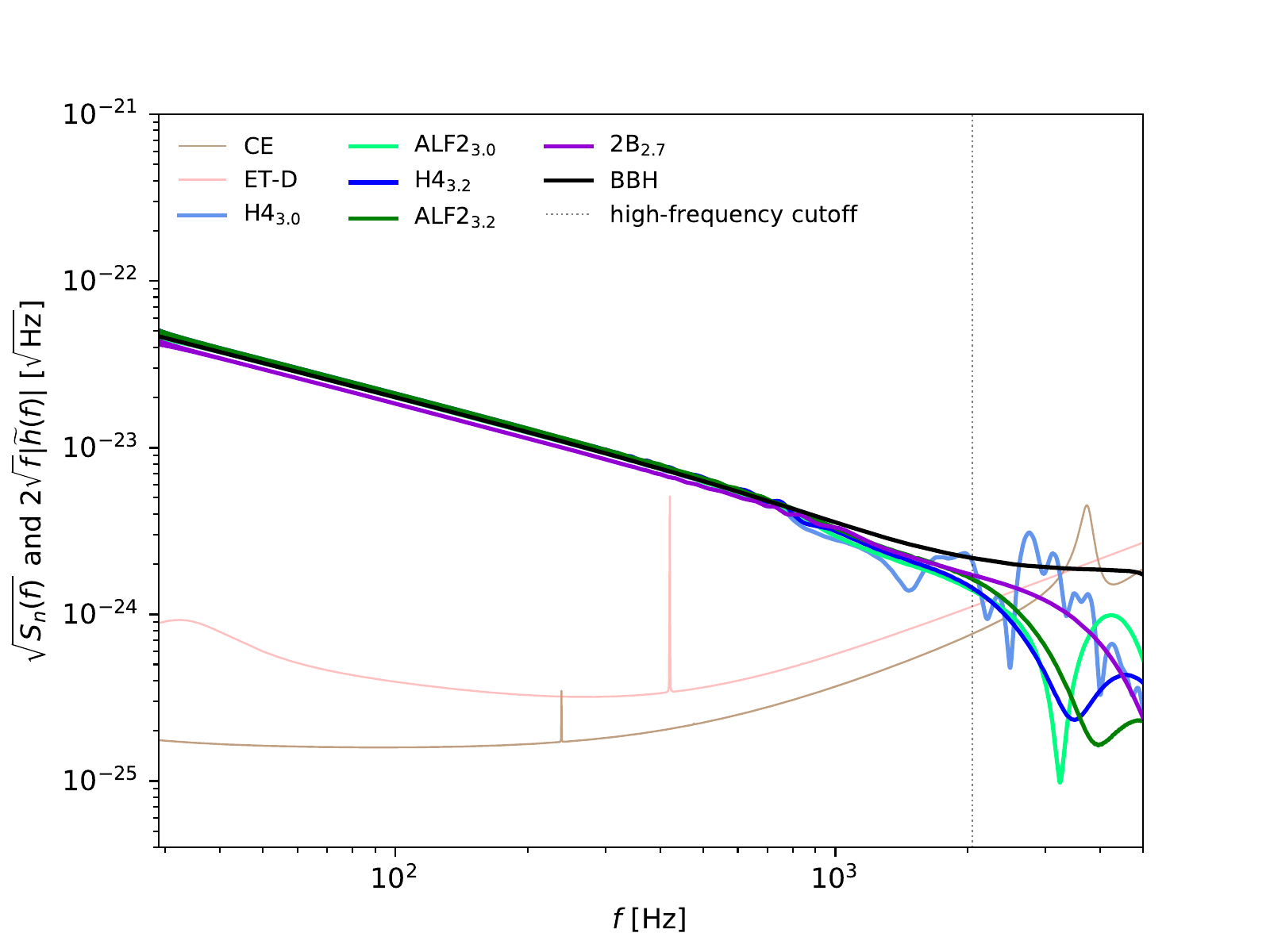}
\caption{The hybrid BNS and TEOBResumS BBH waveforms' amplitude plotted in the frequency domain over some representative detector amplitude noise curves, with the $40$~Mpc injections with the O3- and O4-like LIGO noise curves in the top panel and the $369$~Mpc injections with the 3G noise curves in the bottom panel. The waveforms are scaled to illustrate the SNR accumulation, as discussed in the text. We also mark the $2048$~Hz high-frequency cutoff of our analysis.}
\label{fig:wfs_FD}
\end{figure}

We consider a total of 48 hybrid waveform injections, which are classified into two groups by distance. The first group is injected with a distance of $40$~Mpc (the distance of GW170817) for the O3- and O4-like networks, to give an optimistic result. We use a distance of $369$~Mpc for the 3G network in order to avoid having an excessively large SNR where the results would be even more affected by waveform systematics than we find our current results are (as discussed in Sec.~\ref{sec:systematics}). However, these results will give conservative bounds on how well we can distinguish rarer closer events (redshift effects are still not large at these distances, so the widths of the posteriors will scale with the inverse distance to good accuracy). The second group is injected at a larger distance: $98$~Mpc for the O3- and O4-like networks and $835$~Mpc for the 3G network, since one expects to detect more events at larger distances.

The oddly specific values of distances beyond $40$~Mpc that we consider are due to an initial bug in the injection code that reduced the effective distance of the injection by a factor of $(1+z)^{-1}$ and which we only discovered after performing the parameter estimation runs. Here $z$ is the cosmological redshift at the original injected distance, which we computed using the Planck~2018 TT,TE,EE+lowE+lensing+BAO parameters from Table~2 in Ref.~\cite{Aghanim:2018eyx}. The original planned injected distances were $40$, $100$, $400$~Mpc, and $1$~Gpc, which give the actual injected distances given above (rounded to the nearest integer) after accounting for the bug. Of course, the redshifts of these injections correspond to their original injected distances in the cosmology used. However, the differences in the redshift compared to the Planck~2018 values at the effective distances are not particularly large, at most $15\%$, corresponding to a change in optimal SNR of $2\%$, in the $1$~Gpc case. For comparison, the injected redshift leads to a change of $14\%$ in the SNR, compared to the unredshifted case.

Each group includes the injections of the 5 BNS hybrid waveforms listed in Table~\ref{tab:properties} and 3 BBH waveforms with the corresponding total masses of 2.7, 3.0, and 3.2 $M_{\odot}$ in each of the three detector networks. The injections all have the same right ascension of $2.9082$~h, declination of $62.8505\degree$, inclination angle of $29.9657\degree$, and polarization angle of $45.3047\degree$, which were generated randomly to give an SNR close to the sky-averaged value for the injection GPS time of $1152346754$.

The optimal SNRs of all injections are given in Table~\ref{tab:SNRs}, both from the $31$~Hz low-frequency cutoff used in our parameter estimation analysis (and with the high-frequency cutoff of $2048$~Hz), as well as the fiducial low-frequency cutoffs of the detectors, which are $10$~Hz for all detectors except for ET (since we are using the older CE noise curve), where we use $5$~Hz, as a standard low-frequency cutoff (see, e.g., Refs.~\cite{Meacher:2015rex,Zhao:2017cbb}), though it is possible that the low-frequency cutoff could be as small as $1$~Hz (see, e.g., Refs.~\cite{Punturo:2010zz,Meacher:2015rex,Chan:2018csa}). Since even a $5$~Hz low-frequency cutoff leads to binary neutron star signals that last more than an hour in band, it is necessary to take into account the time dependence of the detector's response due to the rotation of the Earth (see, e.g., Refs.~\cite{Meacher:2015rex,Zhao:2017cbb,Chan:2018csa}). We do this using the method given in Appendix~\ref{app:time_dep_resp}. We find that including the time-dependent detector response decreases the total SNR in ET by at most $1$, due to rounding, with the largest effect for the lowest-mass systems at the closer distance, as expected. The time-dependent response has a larger effect on the SNR in ET from $5$ to $10$~Hz, but even there it only decreases this SNR from $93.5$ to $91.8$ for the $\text{2B}_{2.7}$ system at $369$~Mpc.

Due to the $40$~km long arms of CE, its response to gravitational waves with frequencies $\gtrsim 1$~kHz is frequency-dependent, unlike the usual long-wavelength response that is appropriate for detectors such as Advanced LIGO at the frequencies to which they are sensitive. In particular, the response decreases as the gravitational wavelength is comparable to or smaller than the detector armlength. These effects are studied in Ref.~\cite{Essick:2017}, and found to have a significant effect on sky localization. We have not included this effect in our analysis, due to its computational expense, but we have checked, using the expressions from Ref.~\cite{Essick:2017}, that the frequency-dependent response only leads to a loss of $\sim 0.3$ in the SNR in the two CE detectors for the signals we consider. We leave further checks of the frequency-dependent response (e.g., its effect on matches or Fisher matrix calculations) to future work.

We inject the waveforms with no noise, effectively averaging over noise realizations~\cite{Nissanke:2009kt}. We plot the injections for the smaller of the two distances in the frequency domain in Fig.~\ref{fig:wfs_FD}, comparing them to the detector noise curves. We scale the waveforms showing how the SNR is accumulated as a function of frequency (cf.\ Fig.~1 in Ref.~\cite{TheLIGOScientific:2016pea} and Fig.~5 in Ref.~\cite{Read:2009yp}), since one can write the SNR integral as
\begin{equation}
\rho^2 = \int_0^\infty \frac{[2\sqrt{f}|\tilde{h}(f)|]^2}{S_n(f)}d\ln f,
\end{equation}
[cf.\ Eq.~(1) in Ref.~\cite{TheLIGOScientific:2016pea}], where $\tilde{h}$ is the Fourier transform of the strain $h$ and $S_n$ is the power spectral density of the noise.

Throughout the article, we perform Bayesian parameter estimation using the {IMRPhenomPv2\_NRTidal} model~\cite{Dietrich:2018uni,Dietrich:2017aum,Hannam:2013oca}. IMRPhenomPv2\_NRTidal
augments the GW phase of a precessing point-particle baseline model~\cite{Hannam:2013oca,Khan:2015jqa} 
with the NRTidal-phase description~\cite{Dietrich:2017aum} and tapers the amplitude of the waveform to zero above the merger frequency. It also includes the effects of the stars' spin-induced quadrupole deformations up to next-to-leading order~\cite{Dietrich:2018uni}, parameterizing the EOS dependence of these deformations in terms of the stars' tidal deformabilities using the Love-Q relation~\cite{Yagi:2016bkt}. We use the aligned-spin limit of this model, to simplify the analysis.\footnote{Note that a restriction to aligned spins is not identical to using the aligned-spin version of this model in LALSuite~\cite{LALSuite}, IMRPhenomD\_NRTidal, since this model does not include the contribution from spin-induced deformations, which are important for accurately describing spinning systems~\cite{Harry:2018hke,Dietrich:2018uni,Samajdar:2019ulq}.}
We sample the likelihood using the Markov chain Monte Carlo sampler implemented in the LALInference code~\cite{Veitch:2014wba} in LALSuite~\cite{LALSuite}.

We find that {IMRPhenomPv2\_NRTidal} is able to reproduce the injected waveforms quite well, even though one might be concerned that the absence of the post-merger signal could lead to missing SNR, particularly in the 3G case: If we compute the SNR of the injected waveforms with {IMRPhenomPv2\_NRTidal} waveforms generated using the injected parameters, we find that there is a difference of $\sim 0.01$ between the optimal SNR and the SNR with {IMRPhenomPv2\_NRTidal} for the smaller distance with the O3-like network. The difference rises to $\sim 0.1$ with the O4-like network and the 3G network (both for the $\text{H4}_{3.0}$ system), but seems still negligible.

%----------
\subsection{Post-merger SNRs}
%----------

If we extend the high-frequency cutoff to $5$~kHz, the highest frequency of the noise curves we are using, we find that the post-merger optimal SNRs (above the merger frequency used in the NRTidal model~\cite{Dietrich:2018uni}) are the largest for the $\text{H4}_{3.0}$ system, which has the least compact stars we consider, the longest post-merger signal in the time domain (see Fig.~\ref{fig:NR_GWs}), and the smallest merger frequency, $1.7$ ($1.6$)~kHz for the $40$ ($369$)~Mpc injections. These SNRs are $0.5$, $1.9$, and $2.5$ for the O3-like, O4-like, and 3G networks, respectively, all at the closer distance (i.e., $40$~Mpc for the O3- and O4-like network and $369$~Mpc for the 3G network)---we will not consider the post-merger SNRs at the larger distance---and using the frequency-dependent response for CE. The post-merger optimal SNRs are at least $1.1$ ($1.4$) for the O4-like (3G) network for all the BNS injections expect for the most compact case, $\text{2B}_{2.7}$, where even the 3G network optimal SNR is $0.8$. The largest post-merger optimal SNR for the BBHs is $0.9$ for the $3.2M_\odot$ case. This also has a much higher merger frequency than any of the BNSs, $3.3$~kHz at $369$~Mpc, versus $2.7$~kHz for the $\text{2B}_{2.7}$ system at $40$~Mpc.

If one instead computes the post-merger SNRs with {IMRPhenomPv2\_NRTidal}, one finds that they are notably smaller in the 3G case---the largest decrease in SNR is $0.7$, for the $\text{H4}_{3.0}$ system. With a high-frequency cutoff of $2048$~Hz, the post-merger optimal SNRs are also reduced, in fact becoming zero for the $\text{2B}_{2.7}$ and BBH systems, since the merger frequency is $> 2048$~Hz. The largest decrease to a nonzero value is again for the $\text{H4}_{3.0}$ system, with a decrease of $0.5$, and a further decrease of $0.4$ to the SNR with {IMRPhenomPv2\_NRTidal} (with the $2048$~Hz cutoff).

%%%%%%%%%%%%%%%%%
\section{Results}
\label{sec:results}
%%%%%%%%%%%%%%%%%

\begin{figure}[tb]
\includegraphics[width=0.5\textwidth]{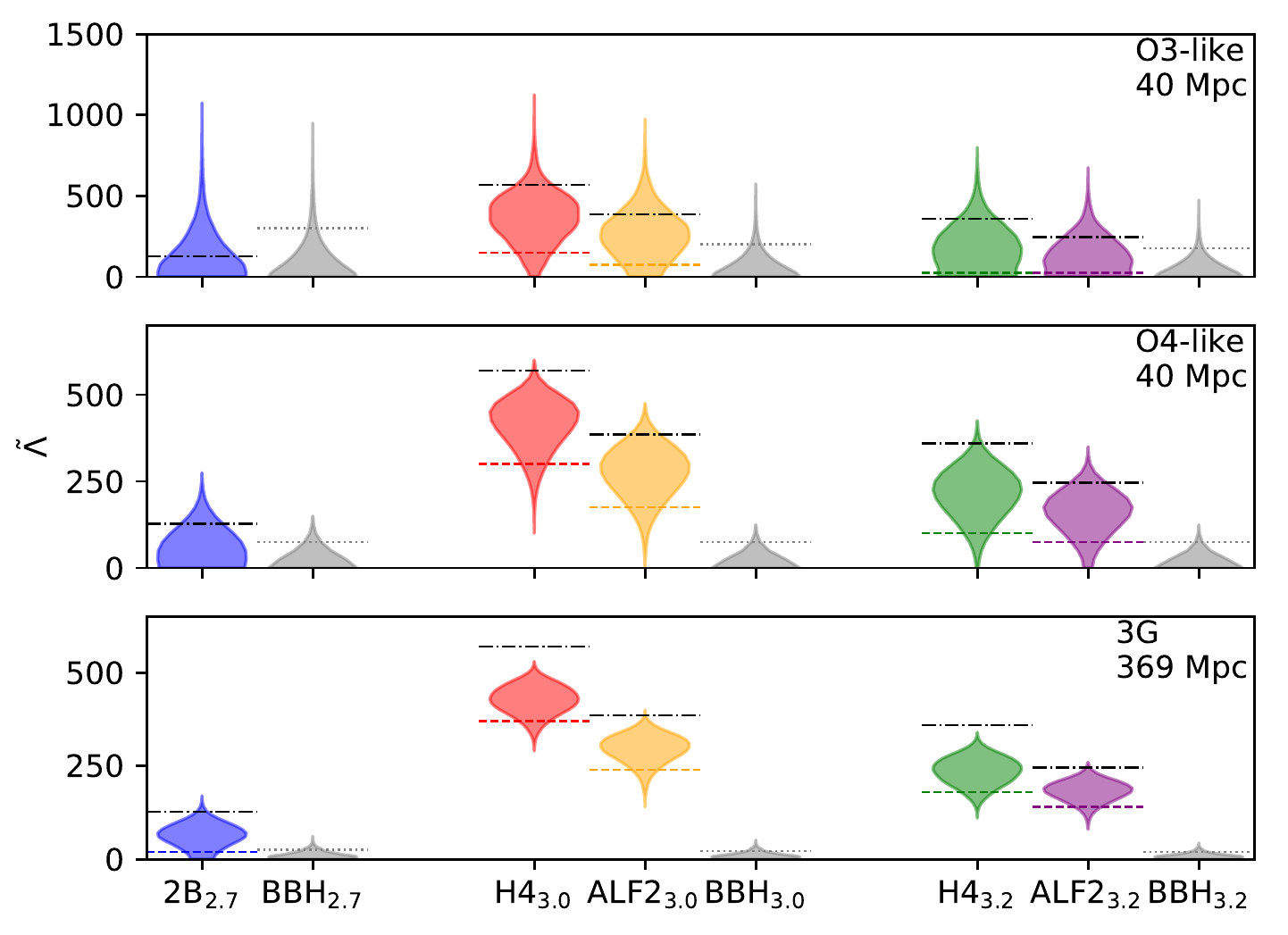}
\includegraphics[width=0.5\textwidth]{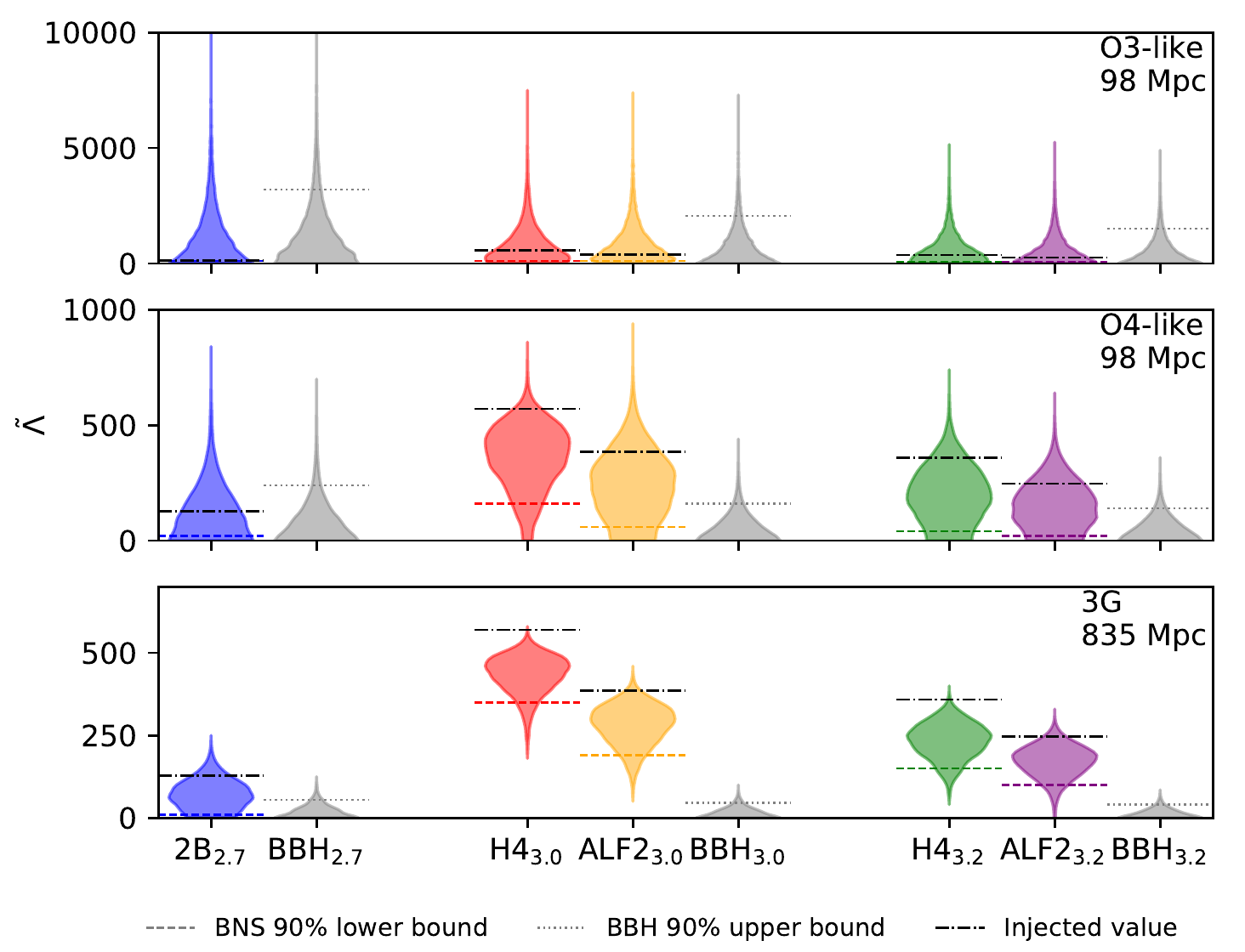}
\caption{Violin plots of the posteriors of $\tilde{\Lambda}$ for the different cases in three detector networks, reweighted to give a flat prior in $\tilde{\Lambda}$. We also show the value of $\tilde{\Lambda}$ for the injection and the $90\%$ credible level lower (upper) bounds on $\tilde{\Lambda}$ for the BNS (BBH) cases.}
\label{fig:lambda_tilde_violin}
\end{figure}

\subsection{Prior choices}

In our analysis, we use the same form of prior probability distributions as in the LIGO-Virgo analyses of GW170817 and GW190425 (see, e.g., Ref.~\cite{GW170817_PE}). Specifically, we assume that
\begin{enumerate}[(i)]
\item the redshifted (detector frame) masses are uniformly distributed;
\item the spins are uniformly distributed in magnitude and direction, with a maximum magnitude of $0.89$ (the high-spin prior used for GW170817 and GW190425)---this is translated to a non-uniform prior on the aligned-spin components we use in this analysis;
\item the individual tidal deformabilities are uniformly distributed and nonnegative;
\item the sources are uniformly distributed in volume (with no redshift corrections applied);
\item the binary's inclination with respect to the observer is uniformly distributed;
\item and the binary's time and phase of coalescence are uniformly distributed.
\end{enumerate}
We use wide enough bounds for all these distributions so that the posterior probability distributions (posteriors, for short) do not have support near the prior limits---the explicit ranges are given in Appendix~\ref{app:prior_ranges}.

\subsection{Posteriors for $\tilde{\Lambda}$ and distinguishability}

\begin{table*}
  \centering
  \caption{\label{tab:credible_level} Maximum credible levels (MCL) at which we can distinguish the $\tilde{\Lambda}$ posteriors for the BNS and BBH injections we consider and Savage-Dickey estimates (SDE) of Bayes factors for the BBH hypothesis $\tilde{\Lambda} = 0$, giving the values as BNS injection v.\ BBH injection. The maximum credible levels are rounded to the nearest percent, except for the values above $99\%$, which are rounded to the nearest tenth of a percent. A dash means that the maximum credible level is $< 80\%$ and the cases where we give lower bounds are those for which we were unable to obtain the true maximum credible level due to a lack of samples in the tails of the distributions. The Bayes factor estimates are given to order-of-magnitude, and the notation for the upper bounds with $<$, $\ll$, $\lll$, and $\llll$ is explained in the text.}
\scalebox{0.95}{

  \begin{tabular}{cccllccllcclrllccllcclrllcclrll}        
    \hline 
    \multirow{4}{*}{Waveform} & & & & & & & \multicolumn{24}{c}{Network \& Distance [Mpc]}\\ 
    \cline{3-31}
     & & \multicolumn{7}{c}{O3-like} & & \multicolumn{9}{c}{O4-like} & & \multicolumn{11}{c}{3G}\\
     \cline{3-9}
     \cline{11-19}
     \cline{21-31}
     & & \multicolumn{3}{c}{$40$} & & \multicolumn{3}{c}{$98$} & & \multicolumn{5}{c}{$40$} & & \multicolumn{3}{c}{$98$} & & \multicolumn{5}{c}{$369$} & & \multicolumn{5}{c}{$835$}\\
     \cline{3-5}
     \cline{7-9}
     \cline{11-15}
     \cline{17-19}
     \cline{21-25}
     \cline{27-31}
     &  & \scriptsize{MCL} & \multicolumn{2}{c}{\scriptsize{SDE}} & & \scriptsize{MCL} & \multicolumn{2}{c}{\scriptsize{SDE}} & & \multicolumn{2}{c}{\scriptsize{MCL}} & & \multicolumn{2}{c}{\scriptsize{SDE}} & &
     \scriptsize{MCL} & \multicolumn{2}{c}{\scriptsize{SDE}} & & \multicolumn{2}{c}{\scriptsize{MCL}} & & \multicolumn{2}{c}{\scriptsize{SDE}} & & \multicolumn{2}{c}{\scriptsize{MCL}} & & \multicolumn{2}{c}{\scriptsize{SDE}}\\     
    \hline 
    ${\rm 2B}_{2.7}$\vphantom{\large\{} & & --  & $10^1$ & \!v.\ $10^1$ & & -- & $10^1$ & \!v.\ $10^1$ & & & -- & & $10^1$ & \!v.\ $10^2$ & & -- & $10^1$ & \!v.\ $10^2$ & & & $89\%$ & & $10^0$ & \!v.\  $10^2$ & & & -- & & $10^1$ & \!v.\ $10^2$ \\
    \hline 
    ${\rm H4}_{3.0}$\vphantom{\large\{} & & $88\%$ & $10^0$\ & \!v.\ $10^1$ & & -- & $10^1$ & \!v.\ $10^1$ & & $>$ & $99\%$ & $\ll$ & $10^{-1}$ & \!v.\  $10^2$ & & $90\%$ & $10^0$ & \!v.\ $10^2$ & & $>$ & $99.4\%$ & $\llll$ & $10^{-1}$ & \!v.\  $10^2$ & & $>$ & $99.6\%$ & $\ll$ & $10^{-1}$ & \!v.\ $10^2$ \\
    ${\rm ALF2}_{3.0}$ & & $81\%$ & $10^0$ & \!v.\  $10^1$ & & -- & $10^1$ & \!v.\ $10^1$ & & & $97\%$ & $<$ & $10^{-1}$ & \!v.\  $10^2$ & & $81\%$ & $10^1$ & \!v.\ $10^2$ & & $>$ & $99.4\%$ & $\lll$ & $10^{-1}$ & \!v.\  $10^2$ & & & $99\%$ & $<$ & $10^{-1}$ & \!v.\ $10^2$ \\
    \hline 
    ${\rm H4}_{3.2}$\vphantom{\large\{} & & -- & $10^1$ & \!v.\  $10^1$ & & -- &  $10^1$ & \!v.\ $10^1$  & & & $97\%$ & $<$ & $10^0$ & \!v.\  $10^2$ & & $80\%$ & $10^1$ & \!v.\ $10^2$ & & $>$ & $99.2\%$ & $\lll$ & $10^{-1}$ & \!v.\  $10^2$ & &  & $99.2\%$ & $\ll$ & $10^0$  & \!v.\ $10^2$ \\
    ${\rm ALF2}_{3.2}$ & & -- & $10^1$ & \!v.\  $10^1$ & & -- & $10^1$ & \!v.\ $10^1$ & & & $95\%$ & & $10^0$ & \!v.\  $10^2$ & & -- & $10^1$ & \!v.\ $10^2$ & & $>$ & $99.2\%$ & $\lll $ & $10^{-1}$ & \!v.\  $10^2$ & &  & $97\%$ & & $10^0$ & \!v.\ $10^2$ \\ 
    \hline 
  \end{tabular}
  }
\end{table*}

The uniform priors on the individual tidal deformabilities lead to a nonuniform prior on the effective tidal deformability $\tilde{\Lambda}$. We thus reweight the $\tilde{\Lambda}$ posterior to have a uniform prior in $\tilde{\Lambda}$ (as in, e.g., Refs.~\cite{GW170817,GW170817_PE,GWTC-1}),\footnote{See Ref.~\cite{Kastaun:2019bxo} for a discussion about caveats concerning
this reweighting procedure; future work may consider the modified reweighting procedure used in the LIGO-Virgo GW190425 analysis~\cite{GW190425}.} obtaining the results given in Fig.~\ref{fig:lambda_tilde_violin}. We compare the $\tilde{\Lambda}$ posteriors for the BNS systems to those of the corresponding BBH systems (instead of comparing both with $0$ directly) in order to reduce the effects of waveform systematics. We compute one-sided credible intervals of $\tilde{\Lambda}$ (upper bounds for the BBH systems and lower bounds for the BNS systems) and consider the maximum credible level at which the BBH upper bound is smaller than the BNS lower bound. We then say that we are able to distinguish the BBH and BNS systems at that credible level.

Since we are considering the tails of the distributions, we estimate the uncertainty in the determination of the quantiles by applying the binomial distribution method given in, e.g., Ref.~\cite{briggs2018a} to obtain the confidence intervals for the quantiles. Here we compute a given quantile $\tilde{\Lambda}_X$ of the reweighted distribution of $\tilde{\Lambda}$ using the Gaussian kernel density estimate (KDE) of the posterior probability density function, and then compute the confidence interval of the quantile of the (unreweighted) samples that has the value $\tilde{\Lambda}_X$. We then reweight the confidence interval to find the estimated uncertainty in the quantile of the reweighted distribution; we truncate the quantiles at $0$ and $1$ when the reweighted confidence interval extends beyond those bounds. The maximum credible levels we quote in Table~\ref{tab:credible_level} include the contribution from the reweighted confidence interval---we multiply the credible and confidence levels together, and take the confidence level to be the larger of $99\%$ and the credible level.

We find that we are not able to distinguish the BNS waveforms from BBH waveforms at the $90\%$ credible level in the O3-like case---the largest credible level at which we can distinguish the posteriors in the O3-like cases is $88\%$, for the ${\rm H4}_{3.0}$ case, the one with the largest $\tilde{\Lambda}$. Moreover, we are not able to distinguish the BNS with the smallest $\tilde{\Lambda}$, ${\rm 2B}_{2.7}$, from a BBH at the $90\%$ credible level in any of the cases we consider---the largest credible level here is $89\%$, in the 3G $369$~Mpc case. However, we are able to distinguish the BNS and BBH posteriors for the H4 and ALF2 waveforms at high credible levels ($95\%$ to $>99\%$) in the O4-like case at $40$~Mpc and at even higher credible levels ($97\%$ to $> 99.6\%$) in the 3G case at both distances. Perhaps surprisingly, the lower bound on the maximum credible level is slightly larger in the 3G $\text{H4}_{3.0}$ case at $835$~Mpc than at $369$~Mpc. However, this is likely explained because the $835$~Mpc case accumulated $20\%$ more samples than the $369$~Mpc case.

%.............
\subsubsection{Bayes factors}
%.............

We can also estimate the Bayes factor in favor of the BBH model using the Savage-Dickey density ratio~\cite{dickey1971}. Here we apply this to the reweighted $\tilde{\Lambda}$, so we compare the Gaussian KDE of the posterior probability density function for the reweighted $\tilde{\Lambda}$ at $\tilde{\Lambda} = 0$ to the flat prior density. We find that the densities we estimate are fairly sensitive to the KDE bandwidth, so we only quote the order of magnitude we obtain, particularly since small differences in the Bayes factor are not very meaningful. We also evaluate the KDE at the smallest value of $\tilde{\Lambda}$ found in the MCMC samples of the posterior probability density, instead of $\tilde{\Lambda} = 0$, to avoid extrapolating the KDE into a region where it is not valid. This is particularly an issue for the cases where it is possible to distinguish the BNS signal from a BBH with high confidence, and the values we report in Table~\ref{tab:credible_level} for such cases are quite conservative upper bounds.

To give a quantitative estimate of approximately how strongly the BBH model would be disfavored in the cases where one can easily distinguish BNSs from BBHs, we extrapolate the tails of the distribution using the skew-normal distribution~\cite{Azzalini}, fixing its parameters by computing the mean, variance, and skew of the KDE (i.e., the method of moments). The skew-normal distribution reproduces the KDEs reasonably well, though since this is purely a phenomenological fit, we only indicate the approximate order of magnitude of the density ratio we obtain from this extrapolation. We do so using $<$, $\ll$, $\lll$, and $\llll$ to indicate extrapolated values with orders of magnitude in the ranges $10^{-1}$, $(10^{-5}, 10^{-2}]$, $(10^{-10}, 10^{-5}]$, and $(10^{-20}, 10^{-15}]$ times the conservative upper limit, respectively. For instance, in the O4-like $\text{H4}_{3.0}$ case at $40$~Mpc, where we quote $\ll 10^{-1}$, the order of magnitude of the extrapolated Savage-Dickey density ratio lies in $(10^{-6}, 10^{-3}]$. The density estimates obtained by extrapolation with the skew-normal distribution are all larger (i.e., more conservative) than those obtained with a pure normal distribution using the same moment method (which is not nearly as good a fit to many of the probability distributions). They are also significantly larger than those obtained by fitting the skew-normal distribution using the maximum likelihood method in SciPy~\cite{Virtanen:2019joe}, which do reproduce the probability distribution better than the method of moments. We quote the results from the method of moments to be more conservative.

We find from the extrapolated Savage-Dickey estimates of the Bayes factors that there is highly significant evidence against the BBH model for BNS injections in the $\text{H4}_{3.0}$ O4-like case at $40$~Mpc, and strong to quite strong evidence against the BBH model for the $\text{H4}_{3.2}$ and $\text{ALF2}_{3.0}$ cases, respectively. In the 3G case, there is extremely strong evidence against the BBH hypothesis for all of the BNS injections at $369$~Mpc, except for the $\text{2B}_{2.7}$ case, where there is fairly strong evidence against the BBH hypothesis, even though we cannot distinguish the posteriors at higher than the $89\%$ credible level. The Bayes factors for the 3G case at $835$~Mpc have almost identical orders of magnitude to those in the O4-like case at $40$~Mpc. However, while the BBH model can be rejected strongly for the BNS injections for which we find the posteriors can be distinguished at large maximum credible levels, the Bayes factor in favor of the BBH hypothesis for BBH injections does not increase in order of magnitude beyond $10^2$ even for the 3G case at $369$~Mpc.

%.............
\subsubsection{Distinguishability of NSBHs}
%.............

Returning to distinguishing the posteriors, we can make a simple test to consider how easy it might be to distinguish these BNS and BBH signals from NSBH signals. For the purposes of this simple test, we assume that the EOS is known exactly and is the same as that used for the BNS injection. We also only consider the O4-like and 3G cases with the H4 and ALF2 EOSs where we can distinguish the BNS and BBH posteriors at a high credible level. We then compute the $\tilde{\Lambda}$ distribution predicted for an NSBH given the EOS and the posteriors on the individual masses, reweight this posterior to the flat prior, and consider the maximum credible level at which this posterior is distinguishable from the measured $\tilde{\Lambda}$ posterior, accounting for sampling uncertainty as above. We consider the cases where both the heavier and lighter neutron stars are replaced by a black hole to be more conservative: The case where the lighter neutron star is replaced by a black hole gives a larger $\tilde{\Lambda}$ than the opposite case, thus making it more difficult to distinguish from a BNS with this method.

We find that in the 3G cases at $369$~Mpc, the BNS (BBH) posteriors are able to be distinguished from the NSBH distributions at greater than the $91\%$ ($99\%$) credible level, with the $\text{ALF2}_{3.2}$ ($\text{H4}_{3.2}$) case giving the smallest credible level and the $\text{H4}_{3.0}$ ($\text{ALF2}_{3.0}$) case giving the largest, of $99.6\%$ ($99.6\%$). In the 3G cases at $835$~Mpc, the maximum credible levels are smaller, at least $85\%$ ($88\%$) in the BNS (BBH) case and at most $98\%$ ($95\%$), with the smallest and largest credible levels occurring for the same cases as at the shorter distance. The maximum credible levels are even smaller in the $40$~Mpc O4-like case, but are still greater than $80\%$ ($86\%$) in the BNS (BBH) case and as large as $96\%$ ($95\%$), again with the same cases giving the smallest and largest values.

The assumption that the EOS is known exactly is likely quite reasonable in the 3G case we consider, since CE Stage 2 will only be operational in $\sim 2044$~\cite{Reitze:2019iox}. However, it is definitely not a good assumption for O4 observations, so a detailed model comparison in that case would likely find that it much more difficult to distinguish these cases. 

\subsection{Posteriors for other parameters}

\begin{figure}[tb]
\includegraphics[width=0.5\textwidth]{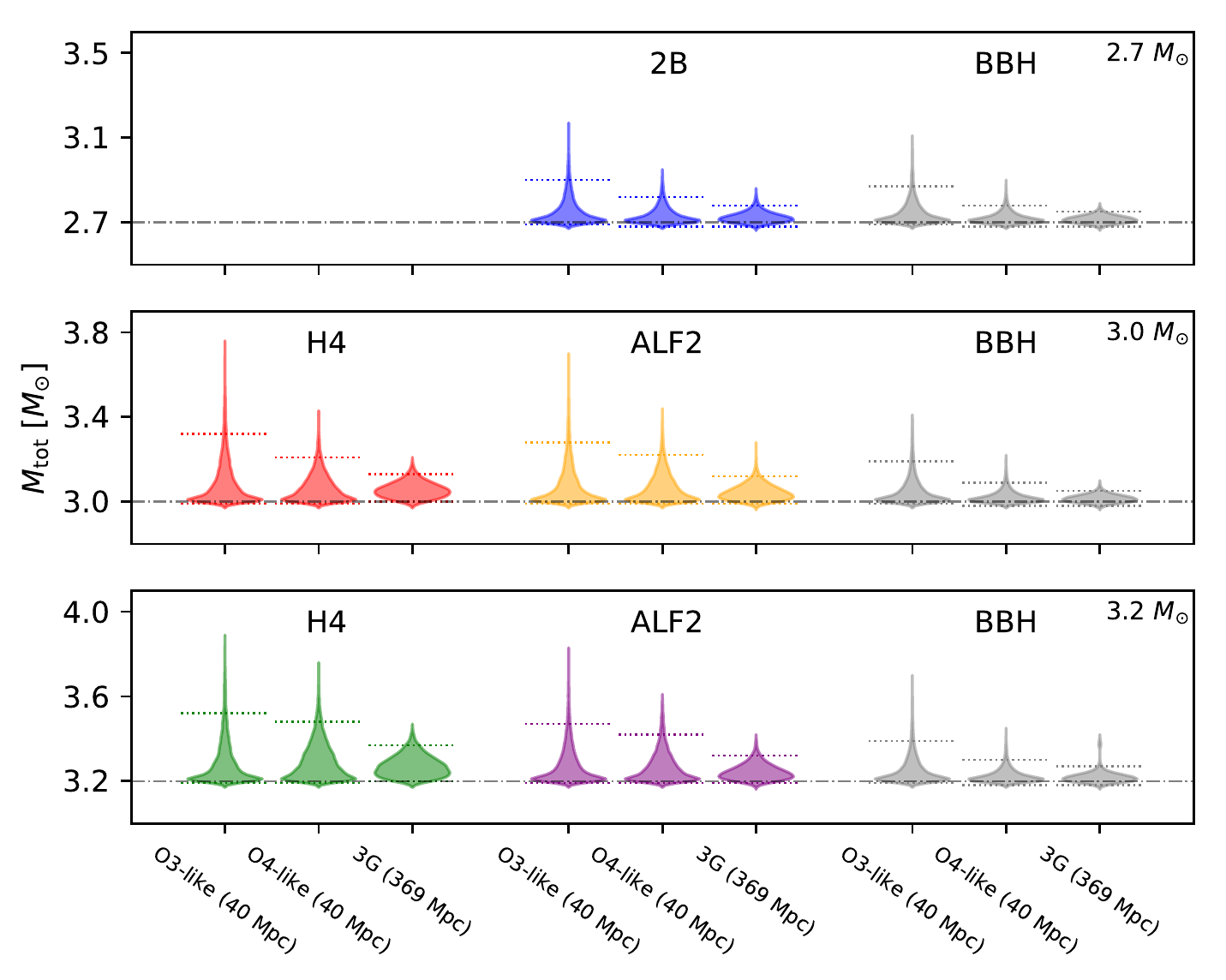}
\includegraphics[width=0.5\textwidth]{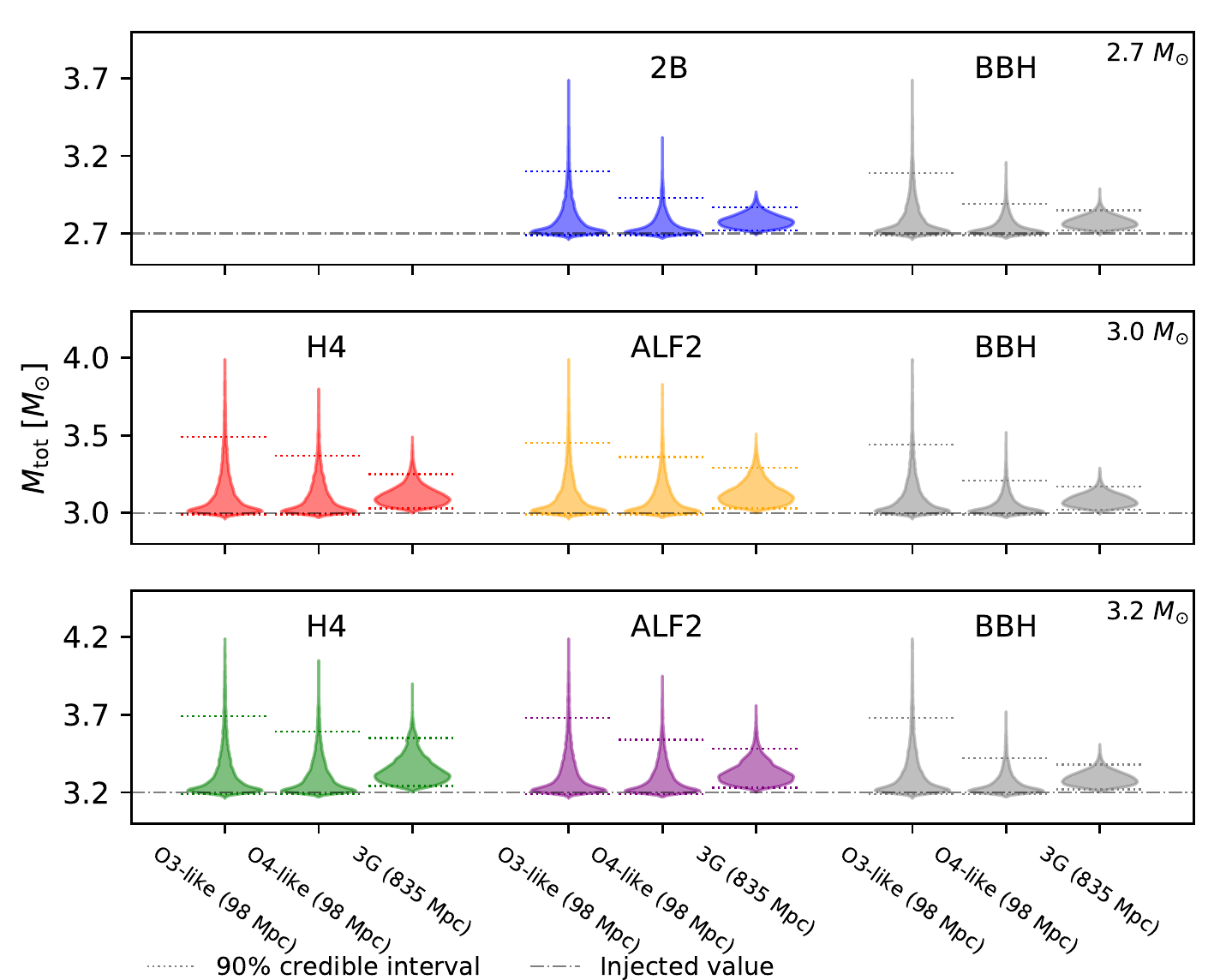}
\caption{Violin plots of the posteriors of the unredshifted (source frame) total mass $M_\mathrm{tot}$ for the different cases in three detector networks. We also show the value of $M_\mathrm{tot}$ for the injection and the $90\%$ credible level interval around the median of $M_\mathrm{tot}$ posterior.}
\label{fig:mass_total_violin}
\end{figure}

\begin{figure}[t]
\includegraphics[width=0.5\textwidth]{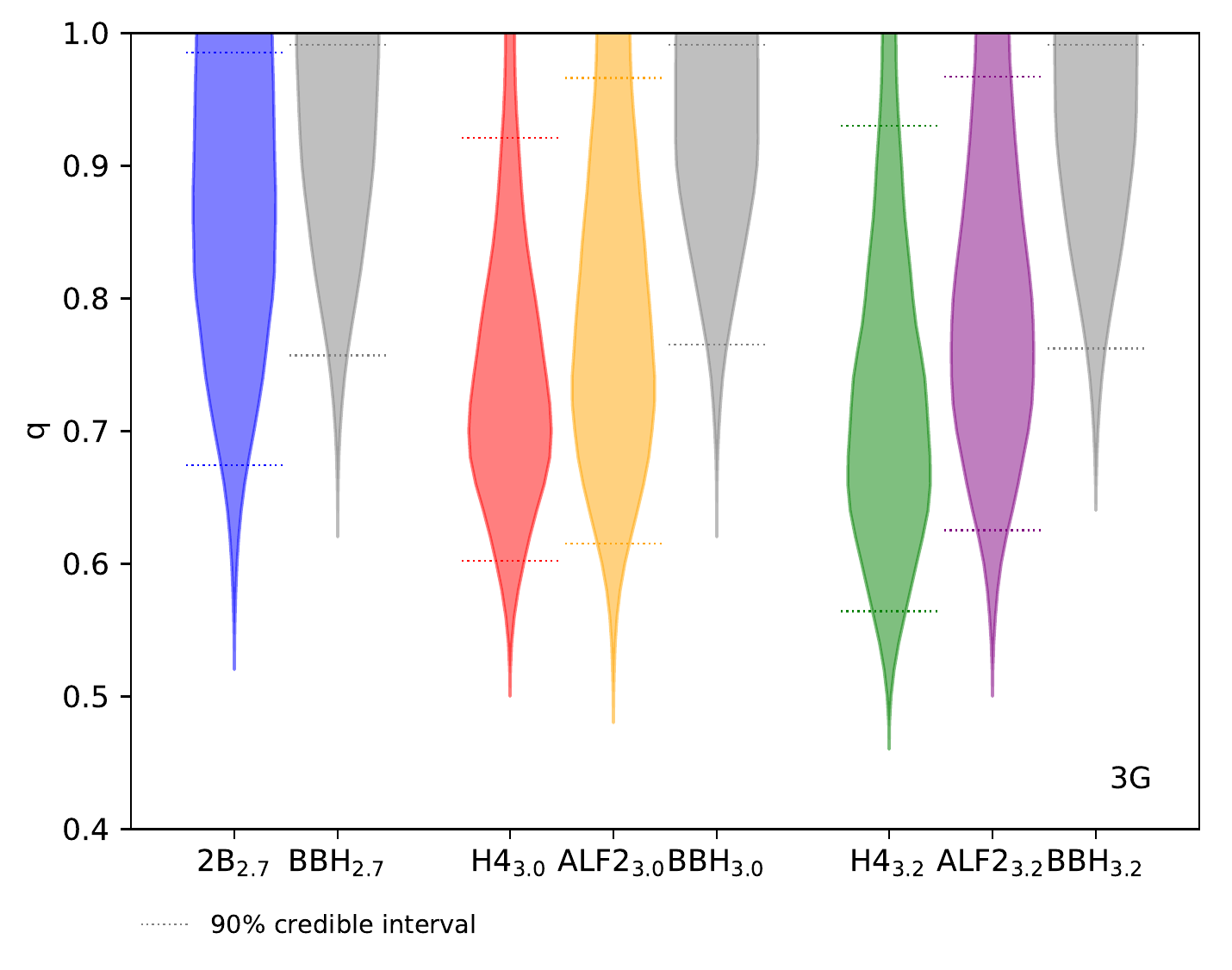}
\caption{Violin plots of the posteriors of the mass ratio $q$ for the different cases in the 3G detector network with an injected distance of $369$~Mpc, showing the $90\%$ credible level intervals around the median. The injected value is $q = 1$ in all cases.}
\label{fig:mass_ratio_violin}
\end{figure}

\begin{figure}[tb]
\includegraphics[width=0.5\textwidth]{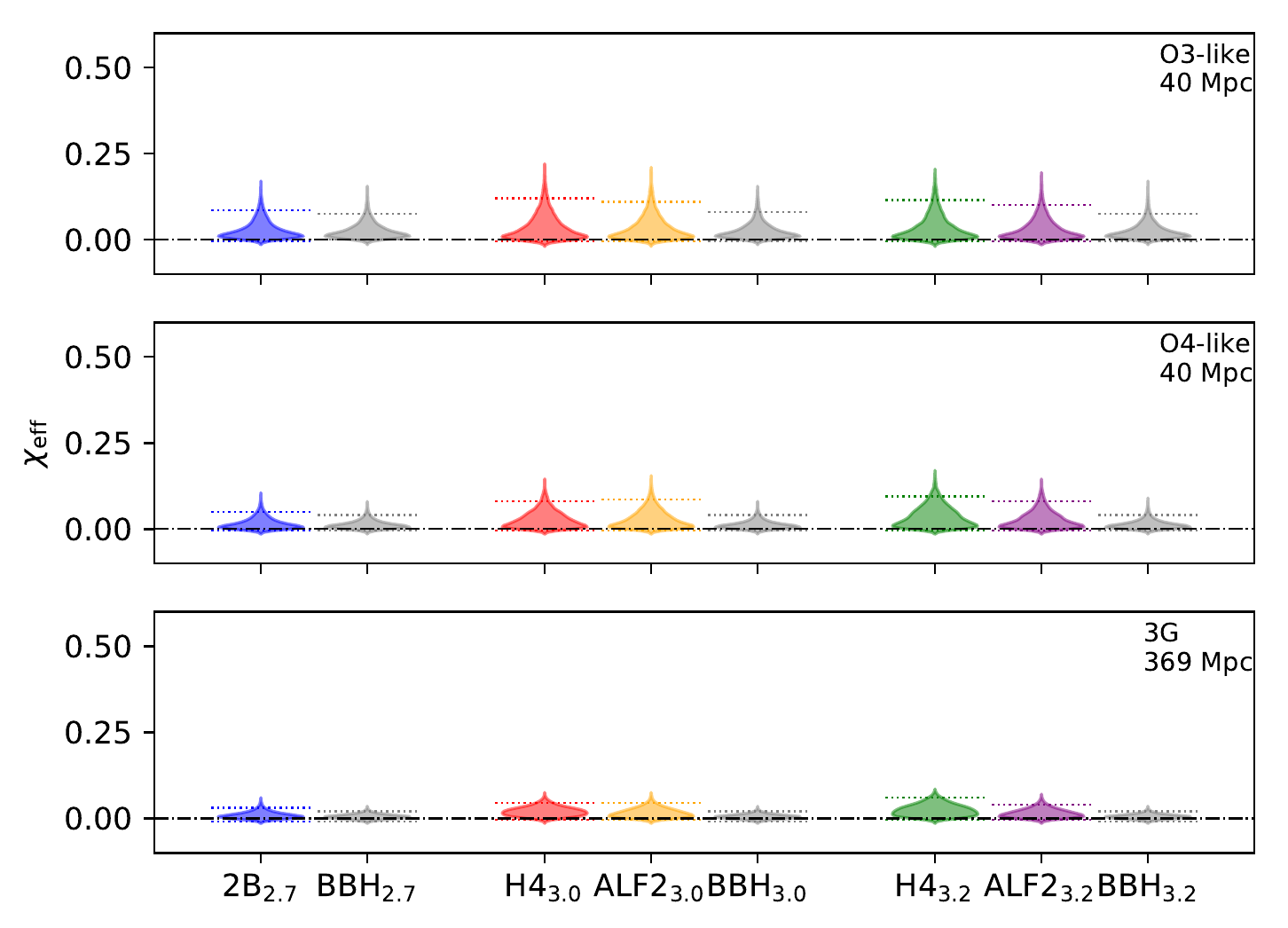}
\includegraphics[width=0.5\textwidth]{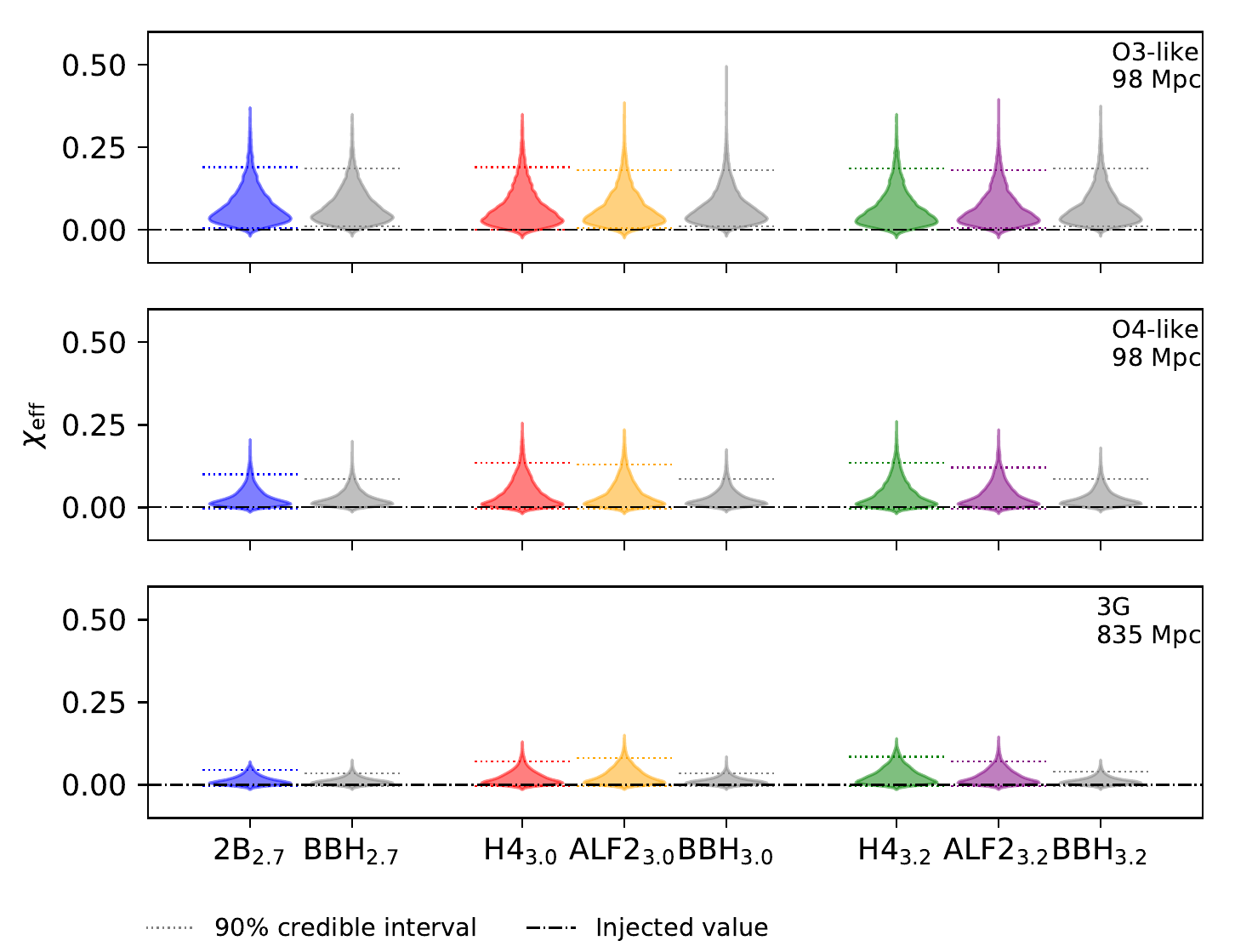}
\caption{Violin plots of the posteriors of the effective spin $\chi_\mathrm{eff}$ for the different cases in three detector networks. We also show the injected value $(\chi_\mathrm{eff} = 0)$ and the $90\%$ credible level interval around the median of the $\chi_\mathrm{eff}$ posterior.}
\label{fig:chi_eff_violin}
\end{figure}

We find that the $90\%$ lower bounds on the unredshifted (source frame) total mass are quite close to the injected values, while the posterior distribution extends to higher total masses---see Fig.~\ref{fig:mass_total_violin}---as expected for an equal-mass injection with a well-determined chirp mass ($90\%$ uncertainty of at most $4\%$, even after converting to the source frame), which then gives a sharp lower bound on the total mass through the relation $M_\text{chirp} = M_\text{tot}\eta^{3/5}$, where the symmetric mass ratio $\eta = M^AM^B/M_\text{tot}^2 \in [0,\nicefrac{1}{4}]$, so that $M_\text{tot} \geq 4^{3/5}M_\text{chirp}$. However, the posteriors are not too broad, with the $90\%$ credible region for the total mass being $< 10\%$ of the injected value. Thus, we would easily be able to identify these systems as potential high mass BNSs.

The mass ratio is estimated far less accurately. Even in the 3G case with a distance of $369$~Mpc, illustrated in Fig.~\ref{fig:mass_ratio_violin}, the $90\%$ credible bound on the mass ratio is not above the $0.8$ approximate bound for a negligible disk mass from Ref.~\cite{Shibata:2019wef}. Moreover, the mass ratio posterior peaks well away from the injected value of $1$ for the H4 and ALF2 BNS cases. This is due to waveform systematics in the IMRPhenomPv2\_NRTidal model, as discussed below.

The effective spin $\chi_\text{eff} := (M^A\chi^A + M^B\chi^B)/M_\text{tot}$ (where $\chi^{A,B}$ are the projections of the dimensionless spins of the individual objects along the binary's orbital angular momentum) is a well-determined spin quantity, and is constrained to be within $\sim 0.1$ or better of the injected value of zero (at the $90\%$ credible level), with negative values strongly disfavored, though the posterior often peaks slightly away from zero, again due to waveform systematics---see Fig.~\ref{fig:chi_eff_violin}. The individual spins are constrained to have $90\%$ upper bounds on their magnitudes of at most $0.33$ (for the secondary of the $\text{2B}_{2.7}$ system at $98$~Mpc), with the primary of the $\text{H4}_{3.0}$ system at $369$~Mpc with the 3G network giving the best constraint, of $0.10$; the constraint on the secondary's spin magnitude in this case is $0.12$. Of course, the portion of the signal below $31$~Hz that we have omitted in this study can help improve the estimation of the mass ratio and spins---see, e.g., Fig.~2 in~\cite{Harry:2018hke}.

The binaries' distances are also estimated reasonably precisely, while the inclination angles are not measured so precisely---we will quote the width of the $90\%$ credible interval to give a measure of the accuracy of these estimates: For the injections at $40$~Mpc, the distances are estimated with an accuracy of $\sim 30\%$ of the injected value for the O3- and O4-like networks, while for the injections at $98$~Mpc, the distance uncertainty is $\sim 50\%$ ($\sim 40\%$) of the injected value for the O3-like (O4-like) network. For the 3G network, the distance uncertainties are $\sim 20\%$ for both distances. For GW170817 (GW190425), the distance was estimated with a fractional accuracy of $60\%$~\cite{GWTC-1} ($90\%$~\cite{GW190425}).
The inclination angle is estimated with an absolute accuracy of $\sim 40^\circ$ ($\sim 50^\circ$), $\sim 30^\circ$ ($\sim 40^\circ$), and $\sim 30^\circ$ ($\sim 30^\circ$) with the O3-like, O4-like, and 3G networks respectively for the closer (further) distance. For comparison, the accuracy of the inclination angle was $\sim 50^\circ$ for GW170817 and unconstrained for GW190425.

We find that these binaries are well localized, with smaller $90\%$ credible regions on the sky than the $16\text{ deg}^2$ final sky localization of GW170817~\cite{GW170817_PE}. This is expected, given that the binaries are observed with SNRs $> 10$ in all detectors in $3$- or $4$-detector networks. Thus, searches for any EM counterpart would be efficient. The size of the sky localization is primarily dependent on the source's distance and the network observing it. The O3-like network gives $90\%$ credible regions of $\sim 5$ and $\sim 10 \text{ deg}^2$ for injections at $40$ and $98$~Mpc, respectively, while the O4-like and 3G networks give $90\%$ credible regions of $\sim 4 \text{ deg}^2$ for all the injections considered. For comparison, the field of view of the Rubin Observatory (formerly known as the Large Synoptic Survey Telescope; under construction) is $9.6\text{ deg}^2$~\cite{Ivezic:2008fe}, while that of the Zwicky Transient Factory is $47\text{ deg}^2$~\cite{ZTF_paper}.

\subsection{Investigation of waveform systematics}
\label{sec:systematics}

Since the $\tilde{\Lambda}$ posteriors in Fig.~\ref{fig:lambda_tilde_violin} peak away from the injected values, as do the mass ratio and effective spin posteriors in Figs.~\ref{fig:mass_ratio_violin} and~\ref{fig:chi_eff_violin}, we consider the difference in matches between the hybrid and {IMRPhenomPv2\_NRTidal} waveforms with the maximum likelihood parameters and the injected parameters, to indicate how large a difference between the waveforms causes this bias in the recovered tidal deformability; see, e.g., Fig.~6 in both Refs.~\cite{Dudi:2018jzn} and~\cite{Samajdar:2018dcx} for a study of biases due to waveform systematics. 
For simplicity, we only vary the masses, spins, and tidal deformabilities and compute the matches using the complex strain ($h_+ - i h_\times$) and the noise curve of one of the detectors in the network. We find that the match between the BNS hybrid and the {IMRPhenomPv2\_NRTidal} waveform with the maximum likelihood masses, spins, and tidal deformabilities is larger than the match with the injected values. However, for the BBH cases, the injected parameters give a slightly larger match than the maximum likelihood parameters, This is expected in cases where there are no significant waveform systematics, since the stochastic sampling used to obtain the maximum likelihood parameters is just obtaining a good approximation to the true maximum likelihood parameters.

We find that the smallest mismatch between the hybrids and the {IMRPhenomPv2\_NRTidal} waveforms with the maximum likelihood parameters is $2\times10^{-5}$, for the ${\rm 2B}_{2.7}$ 3G case with the CE noise curve and the largest mismatch is $7\times10^{-4}$, in the ${\rm H4}_{3.0}$ and ${\rm H4}_{3.2}$ cases with the O4-like Advanced Virgo noise curve. The largest and smallest differences between the matches with {IMRPhenomPv2\_NRTidal} using the maximum likelihood parameters and those using the injected masses, spins, and tidal deformabilities are $0.002$ and $8 \times 10^{-6}$. The largest difference occurs for the same ${\rm H4}_{3.0}$ and ${\rm H4}_{3.2}$ cases with the O4-like Advanced Virgo noise curve that gave the largest mismatch, while the smallest difference occurs for the ${\rm ALF2}_{3.2}$ case with the O3-like Advanced Virgo noise curve. For comparison, the mismatches with detector noise curves due to using a waveform from a lower-resolution simulation to construct the hybrid are $< 4\times10^{-4}$, as illustrated in Appendix~\ref{app:resolution}.\footnote{We have checked that all these small mismatches are computed accurately by comparing with their values with a twice as coarse frequency grid.}

Overall, we find that for systems with larger mismatches, the biases in the recovered parameters, e.g.~Fig.~\ref{fig:mass_ratio_violin}, are largest and that for systems with small mismatches between the BNS hybrid and the {IMRPhenomPv2\_NRTidal} waveform (with the maximum likelihood masses, spins, and tidal deformabilities) the parameter recovery is better. This supports our suggestion that visible biases are caused by waveform systematics. 
Since the mismatches that produce these biases are quite small, well below the maximum mismatches between state-of-the-art EOB models for BNSs (see, e.g., Fig.~21 in~\cite{Nagar:2018zoe}) or between the upgraded {IMRPhenomPv2\_NRTidalv2} model and TEOBResumS--numerical relativity hybrids (see, e.g., Fig.~9 in~\cite{Dietrich:2019kaq}), this indicates that there is significant room for improvement in waveform modeling. Additionally, since we find that the spins make a significant contribution to the maximum likelihood matches in many cases, even though the effective spin is recovered close to the injected value of zero, future improvements to the NRTidal model should likely pay close attention to the modeling of, e.g., spin-induced multipoles.

%%%%%%%%%%%%%%%%%
\section{Conclusions and outlook}
\label{sec:conclusions}
%%%%%%%%%%%%%%%%%

We have investigated how well one can distinguish high-mass equal-mass BNS mergers from BBH mergers using second- and third-generation GW observatories, considering O3-like, O4-like, and 3G networks and using injections of hybridized numerical relativity waveforms. We found that it will be possible to distinguish some reasonably high-mass systems from binary black holes with high confidence. However, it is not possible to distinguish the BNS with the most compact stars we consider from a BBH with high confidence, even with the 3G network. 
Nevertheless, the minimum distance we considered for this system is $369$~Mpc, since this already gives an SNR of $215$, and we did not want to consider significantly higher SNRs due to concerns about waveform systematics---we already found significant waveform systematics at the SNRs considered. It would likely be possible to distinguish this BNS system from a BBH with the 3G network if it merged at a closer distance.
In the future, we will look at this case with improved waveform models, e.g., the IMRPhenomPv2\_NRTidalv2~\cite{Dietrich:2019kaq}, TEOBResumS~\cite{Nagar:2018zoe}, and SEOBNRv4Tsurrogate~\cite{Bohe:2016gbl,Hinderer:2016eia,Steinhoff:2016rfi,Lackey:2018zvw} models. Additionally, it will also be possible to consider binaries with even more compact stars, now that it is possible to construct initial data for such cases~\cite{Tichy:2019ouu}. For cases with very compact stars close to the maximum mass, with $\tilde{\Lambda} \lesssim 10$, a simple scaling of our results suggests that one may only be able to distinguish such BNSs from BBHs using the tidal deformabilitity at quite close distances $\lesssim 50$~Mpc even with the 3G network we consider. However, at these distances, it may be possible to use the post-merger signal, as well, to help discriminate between BNSs and BBHs.

Alternatively, by the time 3G detectors start observing, we will likely have a good estimate of the neutron star maximum mass from inferences of the EOS, from gravitational wave and EM observations---see, e.g., Refs.~\cite{Miller:2019cac,Raaijmakers:2019dks} for constraints on the EOS combining GW170817, NICER, and pulsar mass measurements, Ref.~\cite{Essick:2019ldf} for estimates of the maximum mass from GW170817 constraints on the EOS, and Ref.~\cite{Wysocki:2020myz} for predictions of the accuracy of the maximum mass obtainable by combining together BNS observations---and from possible EM counterpart observations~\cite{Margalit:2017dij,Rezzolla:2017aly,Shibata:2019ctb,Ruiz:2017due}. Thus, if it is possible to constrain the masses of the binary precisely, one can use the maximum neutron star mass to distinguish between BNS, NSBH, and BBH systems, provided that one discounts the possibility of primordial black holes. This would be reasonable if no BBHs are detected at lower masses, where they are easier to distinguish from BNSs. Future work will consider how much the low-frequency part of the signal and higher modes we have omitted here aids in the precise measurement of the individual masses. Additionally, if one has good bounds on the mass ratios for which one expects significant EM counterparts from numerical simulations, EM detections or nondetections could aid in constraining the mass ratio, since the amount of material outside the final black hole, and thus any EM counterparts, is quite sensitive to the mass ratio~\cite{Shibata:2019wef,Coughlin:2018fis,Kiuchi:2019lls}.

Further avenues for exploration include injections of unequal-mass and/or spinning binaries, and the inclusion of spin precession and higher-order modes in the parameter estimation. Additionally, instead of allowing the individual tidal deformabilities to vary independently, as we have done here, it would also be useful to consider model selection comparing BNS, NSBH, and BBH systems where one enforces both neutron stars to have the same EOS in the BNS case, as in, e.g., Ref.~\cite{GW170817_model_comp}. The simple test we made assuming that the EOS is known exactly indicates that it should be possible to distinguish NSBHs from BNSs and BBHs with high confidence in the 3G case. Here one can impose the same EOS in several ways, from phenomenological relations based on expected EOSs, such as the common radius assumption and its extension~\cite{De:2018uhw,Zhao:2018nyf} and the binary-Love relation~\cite{Yagi:2016qmr,Chatziioannou:2018vzf,GW170817_EOS}, to directly sampling in the EOS parameters using a parameterization of the EOS, e.g., the spectral parameterization from Ref.~\cite{Lindblom:2013kra}, as in Refs.~\cite{Carney:2018sdv,GW170817_EOS,GW190425,Wysocki:2020myz}.

We may have already observed the first high-mass BNS merger with GW190425~\cite{GW190425}, though the current detectors were not sensitive enough to determine whether it was indeed a BNS, instead of a BBH or NSBH. While the total mass of GW190425's source is $3.4^{+0.3}_{-0.1}M_\odot$, larger than the total masses considered in this study, if it is an equal-mass system with a total mass of $3.4M_\odot$, it would have $\tilde{\Lambda} = 156$ with the ALF2 EOS and $\tilde{\Lambda} = 73$ with the SLy EOS~\cite{Douchin:2001sv} (a standard soft EOS constructed using the potential method that is consistent with the GW170817 observations~\cite{GW170817_EOS,GWTC-1,GW170817_model_comp}).\footnote{The tidal deformabilities are computed using the IHES EOB code~\cite{Bernuzzi:2014owa,IHES_EOB_code}.} See also the full $\tilde{\Lambda}$ posterior calculated using the GW170817 parameterized EOS results, given in Fig.~14 of~\cite{GW190425}. For comparison, for the $\text{2B}_{2.7}$ system, $\tilde{\Lambda} = 127$. Making a simple scaling of the  $\text{2B}_{2.7}$ results with the 3G network to GW190425's distance of $159^{+69}_{-71}$~Mpc, it seems that it should be possible to distinguish a GW190425-like equal-mass BNS from a BBH at a $> 90\%$ credible level with the 3G network we consider. Of course, direct calculations will be necessary to verify this.

In summary, the prospects for distinguishing high-mass BNSs from BBHs with future GW detector networks are good for the systems we consider, and extrapolating the results for those systems with the 3G network we consider suggests that one will even be able to distinguish BBHs from GW190425-like equal-mass BNS systems, or even more compact BNS systems, if they are sufficiently nearby.

\acknowledgments

It is a pleasure to thank Bernd Br\"ugmann, Katerina Chatziioannou, Reed Essick, Tjonnie Li, B.~S.~Sathyaprakash, Ulrich Sperhake, 
and Wolfgang Tichy for helpful discussions. We also thank Archisman Ghosh for supplying the code used to perform the injections.
A.~C.\ acknowledges support from the Summer Undergraduate Research Exchange programme of the Department of Physics at CUHK and thanks DAMTP for hospitality during his visit.
N.~K.~J.-M.\ acknowledges support from STFC Consolidator Grant No.~ST/L000636/1. Also, this work has received funding from the European Union's Horizon 2020 research and innovation programme under the Marie Sk{\l}odowska-Curie Grant Agreement No.~690904.
T.~D.\ acknowledges support by the European Union's Horizon 2020
research and innovation program under grant agreement No 749145, BNSmergers.
We also acknowledge usage of computer time on the Minerva cluster at
the Max Planck Institute for Gravitational Physics, on
SuperMUC at the LRZ (Munich) under the project number pn56zo, 
and on Cartesius (SURFsarah) under NWO project number 2019.021.
This study used the Python software packages AstroPy~\cite{astropy}, Matplotlib~\cite{hunter2007matplotlib}, NumPy~\cite{oliphant2006guide,van2011numpy}, and SciPy~\cite{Virtanen:2019joe}. This is LIGO document P2000023-v3.

\appendix

\section{Uncertainty of Numerical Relativity Waveforms}
\label{app:resolution}

\begin{table}[t]
  \centering
  \caption{Mismatches between the injected hybrid waveforms and the hybrid waveforms with lower resolution. The lower frequency cutoff for computing the mismatch is 31 Hz.
  \label{tab:match_low_resolution_31} }
  \begin{tabular}{cccccc}        
    \hline 
        \multirow{2}{*}{Waveform} & & \multicolumn{4}{c}{Noise curve}\\
    \cline{3-6}
     & & {flat} & {aLIGO O3-like} & {aLIGO O4-like} & {CE}\\
    \hline 
    ${\rm 2B}_{2.7}$ & & $2.1\times10^{-7}$ & $1.4\times10^{-8}$ & $3.6\times10^{-8}$ & $5.6\times10^{-9}$ \\
    \hline 
    ${\rm H4}_{3.0}$ & & $5.2\times10^{-4}$ & $6.2\times10^{-5}$ & $1.3\times10^{-4}$ & $2.7\times10^{-5}$ \\
    ${\rm ALF2}_{3.0}$ & & $9.3\times10^{-4}$ & $1.6\times10^{-4}$ & $3.9\times10^{-4}$ & $8.5\times10^{-5}$ \\
    \hline 
    ${\rm H4}_{3.2}$  & & $5.7\times10^{-5}$ & $1.1\times10^{-5}$ & $2.8\times10^{-5}$ & $5.3\times10^{-6}$ \\
    ${\rm ALF2}_{3.2}$ & & $6.0\times10^{-4}$ & $1.4\times10^{-4}$ & $3.3\times10^{-4}$ & $7.4\times10^{-5}$ \\ 
    \hline 
  \end{tabular}
\end{table}

\begin{table}[t]
  \centering
  \caption{Mismatches between the injected hybrid waveforms and the hybrid waveforms with lower resolution. The lower frequency cutoff for computing the mismatch is 100 Hz.
  \label{tab:match_low_resolution_100} }
  \begin{tabular}{cccccc}        
    \hline 
        \multirow{2}{*}{Waveform} & & \multicolumn{4}{c}{Noise curve}\\
    \cline{3-6}
     & & {flat} & {aLIGO O3-like} & {aLIGO O4-like} & {CE}\\
    \hline 
    ${\rm 2B}_{2.7}$ & & $1.1\times10^{-6}$ & $3.8\times10^{-8}$ & $8.9\times10^{-8}$ & $3.1\times10^{-8}$ \\
    \hline 
    ${\rm H4}_{3.0}$ & & $2.4\times10^{-3}$ & $1.4\times10^{-4}$ & $2.9\times10^{-4}$ & $1.1\times10^{-4}$ \\
    ${\rm ALF2}_{3.0}$ & & $3.4\times10^{-3}$ & $4.2\times10^{-4}$ & $8.5\times10^{-4}$ & $4.1\times10^{-4}$ \\
    \hline 
    ${\rm H4}_{3.2}$  & & $2.1\times10^{-4}$ & $3.0\times10^{-5}$ & $7.0\times10^{-5}$ & $2.9\times10^{-5}$ \\
    ${\rm ALF2}_{3.2}$ & & $2.1\times10^{-3}$ & $3.6\times10^{-4}$ & $6.8\times10^{-4}$ & $3.5\times10^{-4}$ \\ 
    \hline 
  \end{tabular}
\end{table}

In order to quantify the contribution from the truncation error of the numerical relativity simulation to our injected hybrid waveforms, 
we report in Tables~\ref{tab:match_low_resolution_31} and~\ref{tab:match_low_resolution_100} the mismatches between each of the injected hybrid waveforms and the corresponding hybrid waveform constructed with a lower resolution NR part. The lower resolution simulations have a grid spacing which is about $50\%$ larger than for the highest resolution. We do not vary the settings for the tidal EOB part of the injection and refer the interested reader to Ref.~\cite{Dudi:2018jzn} for additional details. We compute these mismatches with the Advanced LIGO (aLIGO) O3-like and O4-like noise curves and the CE noise curve used in the parameter estimation study (though without any cosmological redshifting), as well as with a flat noise curve, to give a more stringent criterion, without the downweighting of the high-frequency portion given by the detector noise curves. We use a low-frequency cutoff of both $31$~Hz, the same as in the parameter estimation study, and $100$~Hz, to emphasize the higher-frequency portion of the waveform, where tidal effects are more important (see, e.g., Fig.~2 in~\cite{Harry:2018hke}). Even when we start at $100 \rm Hz$, the mismatches between 
the high and low resolution hybrids are small, $< 4\times10^{-3}$ with a flat noise curve and $< 9\times 10^{-4}$ with the detector noise curves.

\section{Including the Earth's rotation in the detector's response}
\label{app:time_dep_resp}

BNS signals can last more than an hour in the CE and ET bands (starting from $5$~Hz), and possibly for days for ET, if its low-frequency cutoff extends down to $1$~Hz---see, e.g., Fig.~2 in Ref.~\cite{Chan:2018csa}---so it is necessary to take the Earth's rotation into account when computing the detector's response. This has been done in Ref.~\cite{Chan:2018csa} in the time domain and in Ref.~\cite{Zhao:2017cbb} in the frequency domain using the stationary phase approximation. Here we show how to account exactly for the effect of the Earth's rotation on the detector's response in the frequency domain, without the Doppler shift (which was found to have a negligible effect on the sky localization in Ref.~\cite{Chan:2018csa}, and will not have a large effect on the SNR), simply by taking the Earth's rotational sidereal angular velocity $\Omega_\oplus$ to be constant, which is true to a very good approximation [fractional errors of $\lesssim 10^{-5}$; see, e.g., Eqs.~(2.11-14) of~\cite{Kaplan:2006nv}]. However, it should be possible to include the effects of the Doppler shift in a similar way (though with further approximations)---see Ref.~\cite{Marsat:2018oam} for similar calculations for LISA.

For a given sky location, we can write the response of a given detector to gravitational waves as a Fourier series in $\Omega_\oplus t$, where $t$ is, e.g., GPS time. Since the response of an interferometric gravitational wave detector is quadratic in its arm vectors, the time dependence of the response has frequencies that are at most twice the Earth's rotational frequency, and thus the Fourier series of the response terminates with the $2\Omega_\oplus t$ terms.

We can therefore write the detector's response (neglecting the Doppler shift) in the time domain as
\<
h(t) = R_+(t)h_+(t) + R_\times(t) h_\times(t),
\?
where
\<
\begin{split}
R_{+,\times}(t) &= a^{+,\times}_0 + a^{+,\times}_{1\text{c}}\cos(\Omega_\oplus t) + a^{+,\times}_{1\text{s}}\sin(\Omega_\oplus t)\\
&\quad + a^{+,\times}_{2\text{c}}\cos(2\Omega_\oplus t) + a^{+,\times}_{2\text{s}}\sin(2\Omega_\oplus t),
\end{split}
\?
and we have not shown the dependence of the Fourier coefficients on the sky location and polarization angle for notational simplicity. In order to calculate the values of the Fourier coefficients, we can evaluate $R_{+,\times}$ at five times, for which we chose $0$, $T_\oplus/8$, $T_\oplus/4$, $T_\oplus/2$, and $3T_\oplus/4$, where $T_\oplus := 2\pi/\Omega_\oplus$ is the Earth's sidereal rotational period, and solve for the coefficients. We denote $R_{+,\times}$ at those times as $R_1$, $R_2$, $R_3$, $R_4$, and $R_5$ respectively (omitting the $+,\times$ labels, for notational simplicity). Then
\<
\begin{split}
R_1 &= a^{+,\times}_0 + a^{+,\times}_{1\text{c}} + a^{+,\times}_{2\text{c}},\\
R_2 &= a^{+,\times}_0 + \frac{\sqrt{2}}{2}(a^{+,\times}_{1\text{c}} + a^{+,\times}_{1\text{s}}) + a^{+,\times}_{2\text{s}},\\
R_3 &= a^{+,\times}_0 + a^{+,\times}_{1\text{s}} - a^{+,\times}_{2\text{c}},\\
R_4 &= a^{+,\times}_0 - a^{+,\times}_{1\text{c}} + a^{+,\times}_{2\text{c}},\\
R_5 &= a^{+,\times}_0 - a^{+,\times}_{1\text{s}} - a^{+,\times}_{2\text{c}}.
\end{split}
\?
Solving the equations above, we can obtain expressions for the coefficients
\<
\begin{split}
a^{+,\times}_0 &= \frac{R_1+R_3+R_4+R_5}{4},\\
a^{+,\times}_{1\text{c}} &= \frac{R_1-R_4}{2},\\
a^{+,\times}_{1\text{s}} &= \frac{R_3-R_5}{2},\\
a^{+,\times}_{2\text{c}} &= \frac{R_1+R_4-R_3-R_5}{4},\\
a^{+,\times}_{2\text{s}} &= R_2 + \frac{(\sqrt{2}-1)(R_4+R_5) - (\sqrt{2}+1)(R_1+R_3)}{4}.
\end{split}
\?
Given a sky location and polarization angle, the values of $R_1$, $R_2$, $R_3$, $R_4$ and $R_5$ can be obtained from standard functions in LALSuite~\cite{LALSuite} or PyCBC~\cite{PyCBC}, e.g., the \verb,antenna_pattern, function in the PyCBC Detector module. We can now easily compute the Fourier transform of $h$ in terms of the Fourier transforms of $h_{+,\times}$, yielding\\
\begin{widetext}
\<\label{eq:htilde_response}
\tilde{h}(f) = \sum_{\alpha\in\{+,\times\}}\left\{a^\alpha_0\tilde{h}_\alpha(f) + \sum_{k\in\{1,2\}}\left[\frac{a^\alpha_{k\text{c}}}{2}[\tilde{h}_\alpha(f - kF_\oplus) + \tilde{h}_\alpha(f + kF_\oplus)] + \frac{a^\alpha_{k\text{s}}}{2i}[\tilde{h}_\alpha(f - kF_\oplus) - \tilde{h}_\alpha(f + kF_\oplus)]\right]\right\},
\?
\end{widetext}
where
\<
\tilde{h}(f) := \int_\R h(t)e^{-2\pi i ft}dt
\?
denotes the Fourier transform, and $F_\oplus := \Omega_\oplus/2\pi \simeq 10^{-5}$~Hz. Since $F_\oplus$ is much smaller than the minimum frequencies it is possible to detect with ground-based detectors ($\gtrsim 1$~Hz), we can approximate the shifts in frequency in Eq.~\eqref{eq:htilde_response} using derivatives of $\tilde{h}_{+.\times}$ (which could be approximated using finite differences of a single frequency-domain waveform), yielding
\begin{widetext}
\<\label{eq:htilde_response_deriv}
\tilde{h}(f) = \sum_{\alpha\in\{+,\times\}}\left\{[a^\alpha_0 + a^\alpha_{1\text{c}} + a^\alpha_{2\text{c}}]\tilde{h}_\alpha(f) + \sum_{k\in\{1,2\}}\left[ika^\alpha_{k\text{s}}F_\oplus\tilde{h}_\alpha'(f) - \frac{k^2a^\alpha_{k\text{c}}F_\oplus^2}{2}\tilde{h}_\alpha''(f)\right] + O[F_\oplus^3\tilde{h}_\alpha'''(f)]\right\}.
\?
\end{widetext}
However, for the current calculations, we used the exact expression in Eq.~\eqref{eq:htilde_response} and leave an exploration of the accuracy of Eq.~\eqref{eq:htilde_response_deriv} for future work.

\section{Prior Bounds}
\label{app:prior_ranges}

\begin{table*}[t]
  \centering
  \caption{The prior bounds used in the analysis. These only depend on the distance and total mass of the injection, except for the cases mentioned in the text.
  \label{tab:prior_bounds} }
  \begin{tabular}{ccccccccccccc}        
    \hline 
        Distance & & $M_{\mathrm{tot}}$ & \hphantom{X}& \multicolumn{9}{c}{Prior bounds}\\
        \cline{5-13}
        [Mpc] & & [$M_{\odot}$] & & $M_{\mathrm{chirp}}$ $[M_{\odot}]$ & & $q$ & & $M^{A,B}$ $[M_{\odot}]$ & & $D_\text{L}$ [Mpc] & & $\Lambda^{A,B}$\\
    \hline 
    \multirow{3}{*}{40} & & $2.7$ & & $1.17$--$1.20$ & & $0.30$--$1$ & & $0.76$--$2.61$ & & $10$--$200$ & & $0$--$3000$\\
     & & $3.0$ & & $1.30$--$1.33$ & & $0.30$--$1$ & & $0.84$--$2.89$ & & $10$--$200$ & & $0$--$3000$\\
     & & $3.2$ & & $1.39$--$1.42$ & & $0.30$--$1$ & & $0.90$--$3.09$ & & $10$--$200$ & & $0$--$3000$\\
    \hline
    \multirow{3}{*}{98} & & $2.7$ & & $1.19$--$1.22$ & & $0.15$--$1$ & & $0.57$--$3.92$ & & $20$--$400$ & & $0$--$20000$\\
     & & $3.0$ & & $1.32$--$1.35$ & & $0.15$--$1$ & & $0.63$--$4.34$ & & $20$--$400$ & & $0$--$15000$\\
     & & $3.2$ & & $1.41$--$1.44$ & & $0.15$--$1$ & & $0.67$--$4.63$ & & $20$--$400$ & & $0$--$15000$\\
    \hline
     \multirow{3}{*}{369} & & $2.7$ & & $1.26$--$1.29$ & & $0.50$--$1$ & & $1.04$--$2.12$ & & $200$--$1000$ & & $0$--$2000$\\
     & & $3.0$ & & $1.40$--$1.43$ & & $0.50$--$1$ & & $1.15$--$2.35$ & & $200$--$1000$ & & $0$--$2000$\\
     & & $3.2$ & & $1.49$--$1.52$ & & $0.50$--$1$ & & $1.22$--$2.50$ & & $200$--$1000$ & & $0$--$2000$\\
    \hline
      \multirow{3}{*}{835} & & $2.7$ & & $1.39$--$1.42$ & & $0.15$--$1$ & & $0.68$--$4.66$ & & $400$--$2000$ & & $0$--$5000$\\
     & & $3.0$ & & $1.55$--$1.58$ & & $0.15$--$1$ & & $0.74$--$5.08$ & & $400$--$2000$ & & $0$--$5000$\\
     & & $3.2$ & & $1.65$--$1.68$ & & $0.15$--$1$ & & $0.79$--$5.40$ & & $400$--$2000$ & & $0$--$5000$\\
    \hline
  \end{tabular}
\end{table*}

In Table~\ref{tab:prior_bounds}, we list the prior bounds used for the chirp mass $M_{\mathrm{chirp}}$, mass ratio $q$, component masses $M^{A,B}$, luminosity distance $D_\text{L}$, and individual tidal deformabilities $\Lambda^{A,B}$. These only depend on the total mass and distance of the system, except for some special cases. For BNS hybrid waveforms with total mass of $3.0 M_{\odot}$ in the O3-like network with a distance of $40$~Mpc, the prior bounds for $q$ are $0.15$--$1$, for $M^{A,B}$ they are $0.62$--$4.27$, and for $\Lambda^{A,B}$ they are $0$--$4000$. The prior bounds for $\Lambda^{A,B}$ for the injected distance of $98$ Mpc shown in Table~\ref{tab:prior_bounds} are only applied to O3-like network. For all the injections with a distance of $98$~Mpc in the O4-like network, the prior bound for $\Lambda^{A,B}$ is $0$--$8000$. For the injected distance of $369$ ($835$)~Mpc, the prior bound of $D_\text{L}$ for the BNS ${\rm 2B}_{2.7}$ and the BBHs with total mass of $3.0 M_{\odot}$ and $3.2 M_{\odot}$ is $100$--$2000$ ($200$--$4000$)~Mpc. For all cases, the aligned spin components are allowed to range from $-0.89$ to $0.89$.

\bibliography{high_mass_bns}

\end{document}